\documentclass[a4paper,11pt]{article}
\pdfoutput=1 
\usepackage{jheppub} 
\usepackage{bm}
\usepackage[T1]{fontenc} 
\usepackage[dvipsnames]{xcolor}
\usepackage{tikz-cd}
\usepackage{comment}
\usepackage{caption}
\usepackage{subcaption}
\usepackage[scr=rsfs]{mathalpha}
\usepackage{bbm}
\usepackage{slashed}
\usepackage{amssymb}
\usepackage{accents}
\usepackage[makeroom]{cancel}
\usepackage{empheq}
\usepackage{leftidx}
\makeatletter
\newsavebox{\@brx}
\newcommand{\llangle}[1][]{\savebox{\@brx}{\(\m@th{#1\langle}\)}%
  \mathopen{\copy\@brx\kern-0.5\wd\@brx\usebox{\@brx}}}
\newcommand{\rrangle}[1][]{\savebox{\@brx}{\(\m@th{#1\rangle}\)}%
  \mathclose{\copy\@brx\kern-0.5\wd\@brx\usebox{\@brx}}}
\makeatother
\usepackage[hyphens]{url} 
\definecolor{linkc}{rgb}{0,0,1}
\usepackage[pagebackref=true,colorlinks=true,linkcolor=linkc,citecolor=linkc,urlcolor=linkc,linktoc=all,pdfusetitle=true]{hyperref}
\renewcommand*{\backref}[1]{}
\renewcommand*{\backrefalt}[4]{{%
		\ifcase #1 
		\or [Cited: pg.~#2.]%
		\else [Cited: pgs. #2.]%
		\fi%
	}}

\usepackage{lineno,enumerate} 
\usepackage[all,2cell]{xy}
\newcommand{\pd}{\partial} 
 
\newcommand{\rd}{\mathrm{d}} 
\newcommand{\bk}{\bm{k}} 
\newcommand{\blm}{\bm{\lambda}} 
\newcommand{\bmu}{\bm{\mu}} 
\newcommand{\hblm}{\hat{\bm{\lambda}}} 
\newcommand{\hbmu}{\hat{\bm{\mu}}} 

\newcommand{\bA}{\bm{A}}
\newcommand{\im}{\mathrm{i}} 
\newcommand{\tsf}[1]{\textsf{#1}}
\newcommand{\ov}[1]{\overline{#1}}
\newcommand{\wt}[1]{\widetilde{#1}}
\newcommand{\wh}[1]{\widehat{#1}}
\usepackage{enumerate}
\usepackage{float}
\usepackage{dsfont}
\usepackage{soul}
\usetikzlibrary{decorations.markings, arrows.meta}
\usepackage[most]{tcolorbox}
\usepackage{longtable}
\usepackage[bottom,hang,flushmargin]{footmisc}
\usepackage{mdframed}

\def\be{ \begin{equation} }
\def\ee{ \end{equation}}
\usepackage[all,2cell]{xy} 

\def\bmod{\mathsf{\,mod\,}}

\makeatletter
\newbox\LT@firstfoot
\def\endfirstfoot{\LT@end@hd@ft\LT@firstfoot}
\newdimen\LT@footdiff
\def\LT@start{%
  \let\LT@start\endgraf
  \endgraf\penalty\z@
  \vskip\LTpre\endgraf
  \LT@footdiff-\ht\LT@foot
  \advance\LT@footdiff\ht\LT@firstfoot
  \dimen@\pagetotal
  \advance\dimen@ \ht\ifvoid\LT@firsthead\LT@head\else\LT@firsthead\fi
  \advance\dimen@ \dp\ifvoid\LT@firsthead\LT@head\else\LT@firsthead\fi
  \advance\dimen@ \ht\ifvoid\LT@firstfoot\LT@foot\else\LT@firstfoot\fi
  \dimen@ii\vfuzz
  \vfuzz\maxdimen
  \setbox\tw@\copy\z@
  \setbox\tw@\vsplit\tw@ to \ht\@arstrutbox
  \setbox\tw@\vbox{\unvbox\tw@}%
  \vfuzz\dimen@ii
  \advance\dimen@ \ht
      \ifdim\ht\@arstrutbox>\ht\tw@\@arstrutbox\else\tw@\fi
  \advance\dimen@\dp
      \ifdim\dp\@arstrutbox>\dp\tw@\@arstrutbox\else\tw@\fi
  \advance\dimen@ -\pagegoal
  \ifdim \dimen@>\z@\vfil\break\fi
  \global\@colroom\@colht
  \ifvoid\LT@firstfoot
    \ifvoid\LT@foot
    \else
      \advance\vsize-\ht\LT@foot
      \global\advance\@colroom-\ht\LT@foot
      \dimen@\pagegoal\advance\dimen@-\ht\LT@foot\pagegoal\dimen@
      \maxdepth\z@
    \fi
  \else
    \advance\vsize-\ht\LT@firstfoot
    \global\advance\@colroom-\ht\LT@firstfoot
    \dimen@\pagegoal\advance\dimen@-\ht\LT@firstfoot\pagegoal\dimen@
    \maxdepth\z@
  \fi
  \ifvoid\LT@firsthead\copy\LT@head\else\box\LT@firsthead\fi\nobreak
  \output{\LT@output}%
}
\def\LT@output{%
  \ifnum\outputpenalty <-\@Mi
    \ifnum\outputpenalty > -\LT@end@pen
      \LT@err{floats and marginpars not allowed in a longtable}\@ehc
    \else
      \setbox\z@\vbox{\unvbox\@cclv}%
      \ifdim \ht\LT@lastfoot>\ht\LT@foot
        \dimen@\pagegoal
        \advance\dimen@-\ht\LT@lastfoot
        \ifdim\dimen@<\ht\z@
          \setbox\@cclv\vbox{\unvbox\z@\copy\LT@foot\vss}%
          \@makecol
          \@outputpage
          \setbox\z@\vbox{\box\LT@head}%
        \fi
      \fi  
      \global\@colroom\@colht
      \global\vsize\@colht   
      \vbox
        {\unvbox\z@\box\ifvoid\LT@lastfoot\LT@foot\else\LT@lastfoot\fi}%
    \fi
  \else
    \ifvoid\LT@firstfoot
      \setbox\@cclv\vbox{\unvbox\@cclv\copy\LT@foot\vss}%
      \@makecol
      \@outputpage
      \global\vsize\@colroom
    \else
      \setbox\@cclv\vbox{\unvbox\@cclv\box\LT@firstfoot\vss}%
      \@makecol
      \@outputpage
      \global\advance\@colroom\LT@footdiff
      \global\vsize\@colroom
    \fi
    \copy\LT@head\nobreak
  \fi
}

\def\C{\mathbb{C}}
\def\CA{{\cal A}}

\def\CC {{\cal C}}

\def\CE {{\cal E}}

\def\CG {{\cal G}}

\def\CM {{\cal M}}
\def\CN {{\cal N}}
\def\CO {{\cal O}}

\def\CO {{\cal O}}

\def\CE {{\cal E}}
\def\CG {{\cal G}}

\def\CS {{\cal S}}

\usepackage{environ}

\NewEnviron{eqsp}{%
  \begin{equation}
    \begin{split}
      \BODY
    \end{split}
  \end{equation}
}
\def\1{\mathds{1}}

\def\IC{\mathbb{C}}

\def\IR{{\mathbb{R}}}

\def\IX{{\mathbb{X}}}

\def\IZ{{\mathds{Z}}}

\def\fe{\mathfrak{e}}

\def\fg{\mathfrak{g}}

\def\fh{\mathfrak{h}}

\def\fl{\mathfrak{l}}

\def\fs{\mathfrak{s}}

\def\fs{\mathfrak{s}}

\def\fu{\mathfrak{u}}

\def\F\IX{\mathfrak{x}}

\def\F\IX{\mathfrak{X}}

\def\hfg{\hat{\mathfrak{g}}}
\def\hfh{\hat{\mathfrak{h}}}

\title{\textsc{A Comment On Topological Degeneracy In Gauged WZW Models}}
\preprint{CERN-TH-2026-153}
\author[a]{Gregory W. Moore,}
\author[b,c,d]{Eliezer Rabinovici,}
\author[a]{Ranveer Kumar Singh}

\affiliation[a]{New High Energy Theory Center and Department of Physics and Astronomy, Rutgers University, 126 Frelinghuysen Rd., Piscataway, NJ 08855, USA}
\affiliation[b]{Racah Institute of Physics,
The Hebrew University,
Jerusalem 9190401,
Israel}
\affiliation[c]{CERN, Theoretical Physics Department,
CH-1211 Geneva 23,
Switzerland}
\affiliation[d]{Department of Theoretical Physics, University of Geneva, 1205 Geneva, Switzerland}
\emailAdd{gwmoore@physics.rutgers.edu}
\emailAdd{eliezer@mail.huji.ac.il}
\emailAdd{ranveer.singh@rutgers.edu}

\abstract{Given a Lie group $G$, a level $k$, and a Lie subgroup $H$ one can construct 2d conformal field theories by either 1.) gauging a nonanomalous $H$ symmetry of the  \tsf{WZW}   model constructed from $(G,k)$ or 2.)  using an algebraic procedure known as the \tsf{GKO} coset construction.  The two models are closely related, but not precisely the same: The gauged \tsf{WZW}   model is identified with the 
corresponding \tsf{GKO} model coupled to a 2d topological field theory.  The topological theory is characterized by a commutative Frobenius algebra derived from the  endomorphisms of an algebra object in a modular tensor category constructed from $(G,H,k)$.  The partition function on 
the torus of the two models differ by a factor of the dimension of this algebra of endomorphisms. Concrete examples are constructed  and some applications to string theory and 2d Yang-Mills coupled to nonanomalous matter are briefly discussed.  This paper is a summary of a longer companion paper. 
\\\\
Date: \today
}

\begin{document}
\maketitle
\flushbottom
\section{Introduction And Conclusion}

This paper addresses the difference between two closely related two-dimensional conformal field theories - the gauged Wess-Zumino-Witten (\tsf{WZW}) model \cite{Bardakci:1987ee,Altschuler:1987zb,rabinovici1988aspects, Gawedzki:1988nj} and the coset construction of Goddard, Kent, and Olive (\tsf{GKO}) \cite{Goddard:1984vk,Goddard:1986ee}.  In the literature, the two theories are often conflated. In this paper we demonstrate that they differ by topological degrees of freedom and we describe the 
relevant 2d Topological Field Theory (\tsf{TFT}) that governs these topological degrees of freedom.

Given a finite-dimensional Lie group $G$ and a ``level'' $k \in H^4(BG,\IZ)$,
where $BG$ is the classifying space of $G$,  one can define a two-dimensional conformal field theory, the $\mr{WZW}$ model. We denote the theory by 
$\mr{WZW}(G;k)$. The $\mr{WZW}$ model is best understood for the case when $G$ is compact, 
in which case    the theory is a rational \tsf{CFT}. We henceforth assume $G$ is compact until section 
\ref{sec:implications} where we comment on the noncompact case, which is particularly relevant in applications to string theory. In this paper we only consider the ``diagonal'' combination of left- and right-movers.   Let $H$ be a compact Lie subgroup of $G$.
Then, there is a nonanomalous $H$-symmetry of $\mr{WZW}(G;k)$,  where $H$ acts  on the \tsf{WZW} field with target manifold $G$ by conjugation \cite{Bardakci:1987ee,Altschuler:1987zb,rabinovici1988aspects, Gawedzki:1988nj}.
We denote the theory with this $H$-action gauged, and with no kinetic term for the $H$-gauge fields, by  $\mr{WZW}(G,H;k)$. This theory is also a rational conformal field theory.

The chiral operator content of $\mr{WZW}(G;k)$ can be constructed from a vertex operator algebra (\tsf{VOA}).
(See \cite{FLM1988,frenkel1993axiomatic,lepowsky2012introduction} for basic material in vertex operator algebra theory.)
We denote this \tsf{VOA}  by $V(G;k)$. 
In the case where $G$ is connected, simply connected and semisimple one forms the affine Lie algebra for the  Lie algebra $\fg = \mr{Lie}(G)$ with level $\bk$. 
\footnote{We can embed $H^4(BG;\IZ) \rightarrow H^4(BG;\IR)$. Then there is a canonical isomorphism of $H^4(BG;\IR)$ with $\mr{Ad}$-invariant quadratic forms on $\fg$. We take $\bk$ to be the image of $k$ and identify it with the level of the affine Lie algebra. }
Then, the basic representation $L_{\fg}(\bk,0)$ of the affine Lie algebra admits the structure of a vertex operator algebra; it can then be identified with  $V(G;k)$. When $G$ is a quotient of a connected, simply connected  group $\tilde G$ the vertex operator $V(G;k)$ is derived from $V(\tilde G;k)$ by adjoining suitable chiral fields with integer spin and no common braiding \cite{Moore:1988ss,Moore:1989yh}.

The full operator content of $\mr{WZW}(G;k)$ is obtained from the state-operator correspondence. The space of states on a circle is the sum over the (diagonal product of) irreducible unitary highest weight modules for $V(G;k)$. Now, because $H$ is a subgroup of $G$ there is a sub-conformal field theory 
$\mr{WZW}(H;\tilde k)$ of $\mr{WZW}(G;k)$ where $\tilde k$ is the pullback of $k$ under 
$H^4(BH;\IZ) \to H^4(BG;\IZ)$. We \emph{define} the \tsf{GKO} vertex operator algebra 
$V(G,H;k)$ to be the commutant of $V(H;\tilde k)$ in $V(G;k)$. See section \ref{sec:GKO_review} for details. It is expected - and proved in a large number of cases - that $V(G,H;k)$ is a rational vertex operator algebra.
\footnote{It has been proved that parafermionic \tsf{VOA} $V(G,H;k)$, with $H$ a maximal torus subgroup of $G$, are $C_2$-cofinite \cite{ALY1,DW2}. The rationality of parafermionic \tsf{VOA} obtained from $\fs\fl_2$ and $\fs\fl_n$ has been proved in \cite{ALY2,JL1,JL2}. } 
One can therefore form the rational conformal field theory whose space of states on a circle 
is the diagonal sum over the irreducible modules of $V(G,H;k)$. We \emph{define} this theory to 
be the \tsf{GKO} conformal field theory and denote it by $\mr{GKO}(G,H;k)$. In the case where $G,H$ are connected, simply connected, and semisimple we also denote the \tsf{GKO} vertex operator algebra by $V(\fg,\fh;k)$ and the \tsf{GKO} theory 
by  $\mr{GKO}(\fg,\fh;k)$.

One of the key results of the paper \cite{Goddard:1984vk} is that the Virasoro charge is given by 
\be\label{eq:GKO-CentralCharge}
c( \mr{GKO}(G,H;k))  = c(\mr{WZW}(G;k))  - c(\mr{WZW}(H;\tilde k) )
\ee
It was, moreover, shown in the papers \cite{Bardakci:1987ee,Altschuler:1987zb,rabinovici1988aspects,Gawedzki:1988nj} 
that the operator content of  $\mr{WZW}(G,H;k)$ is closely related to that of $\mr{GKO}(\fg,\fh;k)$. 
In particular, the formula for the central charge \eqref{eq:GKO-CentralCharge} was recovered. 
Moreover,   the list (without degeneracy) of Virasoro conformal weights that appear in the two theories is identical. 
However, the comparison of the theories produced a puzzle. References \cite{Bardakci:1987ee,Altschuler:1987zb,rabinovici1988aspects} studied the above theories, and in particular the construction of the unitary series of $c<1$ \tsf{CFT}s by gauging subgroups of the $\mr{O}(N)$ symmetry of a modular invariant system of free fermions. The construction of the partition functions in \cite{Bardakci:1987ee,Altschuler:1987zb,rabinovici1988aspects} indicated that the theories so constructed have a nontrivial, but finite, vacuum degeneracy. 
By ``vacuum degeneracy'' we mean the dimension of the space of local  fields of conformal dimension zero in the Hilbert space on a circle. Some authors will declare a ``conformal field theory'' to have a one-dimensional vacuum. We do not do that. We consider the tensor product of a conformal field theory and a topological field theory to be another conformal field theory. An explanation of 
the vacuum degeneracy observed in \cite{Bardakci:1987ee,Altschuler:1987zb,rabinovici1988aspects} is an important open problem and the resolution might have applications in string theory. The existence of a degeneracy in some cases has also recently been discussed in \cite{Komargodski:2020mxz,Cordova:2023jip}.

In this paper, we argue that the vacuum degeneracy observed in \cite{Bardakci:1987ee,Altschuler:1987zb} is a generic feature. We will show that  the operator content of $\mr{WZW}(G,H;k)$ and $\mr{GKO}(G,H;k)$ differ by an interesting topological degeneracy.
We argue that a 2d \tsf{TFT} must be ``coupled'' to $\mr{GKO}(G,H;k)$ to produce 
the theory $\mr{WZW}(G,H;k)$. 
Our main tool is the study of the space of states on the circle and our sharpest result pertains to the case that $H$ and $G$ are connected, simply connected, and semisimple, and, moreover 
satisfy a condition of ``$Z$-regularity'' explained below. An important equation
\emph{en route} to our main result is equation \eqref{eq:GmodH-WZW-space}, but some care is needed to identify the topological degrees implicit in \eqref{eq:GmodH-WZW-space}.

Some insight into the topological degrees of freedom can be gleaned by recalling that for the case $H=G$ the gauged \tsf{WZW} model $\mr{WZW}(H,H;k)$ is a 2d \tsf{TFT}  \cite{SPIEGELGLAS199036,Spiegelglas:1991uc,Spiegelglas:1992jg,Witten:1991mm,Witten:1993xi} and the topological degrees of freedom alluded to above constitute the entire theory. 
Indeed, identifying a 2d oriented topological field theory with a commutative Frobenius algebra,
the theory is just that defined by the \textit{Verlinde algebra} $V_{k}(H)$ of $\mr{WZW}(H;k)$.
\footnote{The topological field theory $\mr{WZW}(H,H;k)$ is important to the Freed-Hopkins-Teleman theorem \cite{Freed:2003qx,Freed:2007wja,Freed:2009qp}. For a physical interpretation of the Freed-Hopkins-Teleman theorem see 
\cite{Moore:2003vf}. For a recent review of topological field theory that explains some results on 2d \tsf{TFT} relevant to this paper see, for example,   \cite{Moore:2025tmt}. See also \cite{Schafer-Nameki:2003nzb} for an extension of \cite{Freed:2007wja,Moore:2003vf} to supersymmetric coset models.}
In this case the \tsf{GKO} model $\mr{GKO}(H,H;k)$ is the trivial theory: The operator algebra consists entirely of the unit operator; the space of states is one-dimensional. 
Turning now to the case where $H$ is a proper subgroup of $G$,   some of the fields in the $G$-theory are just maps into $H$ and therefore we may guess that by gauging $H$ the topological degrees of freedom we seek will be related to the 2d \tsf{TFT} defined by the Verlinde algebra $V_{\tilde k}(H)$.  Under some conditions this will indeed turn out to be the case, as shown in section \ref{sec:equivariant_H/H}. Close inspection shows that there is the interesting   added wrinkle that the 2d \tsf{TFT} we seek is actually a $Z$-equivariant extension of  $V_{\tilde k}(H)$, where 
\be\label{eq:defZ}
Z :=Z(G)\cap Z(H)~,
\ee
is the common center of $G$ and $H$.  The equivariant extension of the \tsf{TFT} $V_{\tilde k}(H)$ by $Z$ is explained in \cite{Moore:2006dw}. Its appearance here is related to the fact that the gauge action of $Z$ on the \tsf{WZW} fields is ineffective. The relevance of this ineffective group action to the phenomenon of \emph{decomposition} has been noted and discussed in \cite{Pantev:2022pbf}.

The interpretation of the topological degrees of freedom in terms of the $Z$-equivariant extension of $V_{\tilde k}(H)$ only holds under certain technical conditions. (These conditions are explained in section \ref{sec:equivariant_H/H}). We are aiming to find a statement valid in the general case of an arbitrary pair of compact Lie groups $H < G$. In order to state a result susceptible of applying at such a level of generality we 
must delve into the mathematics of modular tensor categories. Readers who wish to see examples, such as the unitary minimal models and parafermions, before taking the plunge can skip to section \ref{subsec:Examples} and section \ref{sec:parafermions}. Using the results of \cite{Frohlich:2003hm,Frohlich:2003hg} the modular tensor category (\tsf{MTC}) of the \tsf{GKO} 
theory can be constructed as the category of modules of an algebra object $B$ in the 
\tsf{MTC} of the $(G_k\times H_{-\tilde{k}})$-Chern-Simons theory. The precise definition of $B$ 
is given in   \eqref{eq:B_def_affine}. In the terminology commonly found in the physics literature $B$ is a nonabelian anyon and one ``condenses'' this anyon to obtain the \tsf{MTC} of the \tsf{GKO} theory \cite{bais2009condensate,kong2014anyon}. For some interesting recent works using 
this concept to discuss new dualities in 2d \tsf{CFT} see, for examples, \cite{Cordova:2023jip,Cordova:2024goh,Cordova:2025zkz}.

The representation theory of the \tsf{GKO} vertex operator algebra $V(G,H;k)$ is rather subtle. 
See section \ref{sec:GKO_review} for a more detailed discussion. Briefly,  one defines \emph{branching representations} of $V(G,H;k)$ by considering the 
isotypical decomposition of irreducible representations of $V(G;k)$ in terms of irreducible representations 
of $V(H;\tilde k)$. Denoting the irreducible highest weight representations of $V(G;k)$ by   $L_{\fg}(\bk,\hblm)$, 
with $\hblm$ labeling the distinct representations, and similarly those of 
$V(H;\tilde k)$ by $L_{\fh}(\wt{\bk}, \hbmu)$, the branching representations of 
$V(G,H;k)$ are defined by the isotypical decomposition: 
\be 
L_{\fg}(\bk, \hblm) = \bigoplus_{\hbmu} L_{\fg/\fh}(\hblm,\hbmu) \otimes L_{\fh}(\widetilde \bk, \hbmu)~.  
\ee
One must use caution when working with the branching representations $L_{\fg/\fh}(\hblm,\hbmu) $: 

\begin{enumerate} 

\item There are \emph{selection rules} that tell us for which pairs $(\hblm,\hbmu)$ the representations are nonzero.  

\item It can happen that different pairs $(\hblm,\hbmu)$ and $(\hblm',\hbmu')$ give isomorphic 
representations $L_{\fg/\fh}(\hblm,\hbmu) \cong L_{\fg/\fh}(\hblm',\hbmu')$. This is the phenomenon of 
\emph{field identification}.

\item It can also happen that $L_{\fg/\fh}(\hblm,\hbmu)$ are not irreducible. Resolving the branching representation into irreducible subrepresentations is sometimes referred to as \emph{fixed point resolution}.  

\item It is not \emph{a priori} obvious if $L_{\fg/\fh}(\hblm,\hbmu)$ (or their subrepresentations) generates the full category of $V(G,H;k)$ modules.

\end{enumerate}

The above questions are implicitly resolved by the results of \cite{Frohlich:2003hm,Frohlich:2003hg}. However, translating the category-theoretic result into explicit answers to the above issues is, in general, not straightforward. There is one 
important case where the answer is known. Consider the case when $G,H$ are compact, connected, simply connected and semisimple.
There is an action of $Z$ on the pairs $(\hblm,\hbmu)$ which was used in \cite{Moore:1989yh,gepner1989field,Schellekens:1989uf} to explain selection rules and field 
identifications. See equation \eqref{eq:sel_rules_JJ'} below for selection rules and 
equation \eqref{eq:FieldIdent} for field identifications.
We define  a \tsf{GKO} vertex operator algebra $V(G,H;k)$ to be \emph{$Z$-regular} if 

\begin{enumerate}
   
\item The branching representations are irreducible and generate the category of modules for $V(G,H;k)$. 

\item The selection rule stated in  \eqref{eq:sel_rules_JJ'} is replaced by the statement 
that the character $\chi_{(\hblm,\hbmu)}$ is trivial \textit{if and only if} the module $L_{\fg/\fh}(\hblm,\hbmu)$ is 
nonzero. 

\item All field identifications are obtained from the $Z$-action on the set $\cal{E}$ of pairs $(\hblm,\hbmu)$ for which $L_{\fg/\fh}(\hblm,\hbmu)$ is nonzero. 
\item The $Z$-action on $\cal{E}$ is fixed-point-free. 

\end{enumerate}

There are infinitely many examples of \tsf{GKO} theories that are $Z$-regular and the main 
result in this paper, Theorem \ref{thm:main_thm}, applies to the $Z$-regular case.  
The theorem states that 
%
%
%
\begin{enumerate}

\item\label{it:GKO-WZW1} The  state space on the circle of $\mr{WZW}(G,H;k)$ can be written (see equation \eqref{eq:G/H_space_RepS})  as a 
direct sum over irreducible modules $W^r$ of  $V(G,H;k)$ in the form 
\begin{eqsp}\label{eq:G/H_space_RepS}
    \mathscr{H}^{\mr{WZW}}_{G/H}\cong \bigoplus_{r\in \mr{Irrep}(V(G,H;k))}  \left(W^r \otimes \wt{W}^r\right)\otimes D_r ~.
\end{eqsp}
where $D_r$ are finite-dimensional degeneracy spaces. 
Denoting the vacuum module  $V(G,H;k)$ of the \tsf{GKO} theory  by $r=1$ we have   $D_1 \cong \mr{End}(B)$. 

\item  \label{it:GKO-WZW2}   $\mr{End}(B)$, which can be identified with the vector space of states of conformal dimension zero,  is endowed with a 
commutative Frobenius algebra structure and can thus  be  identified with a 2d \tsf{TFT}.

\item \label{it:GKO-WZW3} Moreover, for all $r$ the $D_r$ are  free rank-one modules for $\mr{End}(B)$. 
 Thus, each module $D_r$ is isomorphic to $\mr{End}(B)$ as a vector space, 
 although for general $r$ we have not found a natural algebra structure on $D_r$ and it is 
 not obvious that one should exist. 

\end{enumerate} 

It is not true that the \tsf{WZW} theory is the tensor product of the \tsf{GKO} theory with a 2d \tsf{TFT} determined by $\mr{End}(B)$. 
Rather the \tsf{GKO} theory and the 2d \tsf{TFT} are ``intertwined.'' We leave as an open problem a deeper understanding of this intertwining. Nevertheless, we will informally summarize the result in the equation: 
\begin{eqsp}\label{eq:WZW=GKOEndB}
    \mr{WZW}(G,H;k)\cong \mr{GKO}(G,H;k) \, \widetilde{\otimes } \, \mr{End}(B)~,
\end{eqsp}
where the tilde in the expression  $\widetilde{\otimes }$ is meant to remind us that the theory on the right hand side of \eqref{eq:WZW=GKOEndB} is not a standard tensor product of theories. The 
definition of $\widetilde{\otimes }$ is given by the properties \autoref{it:GKO-WZW1}, \autoref{it:GKO-WZW2} and \autoref{it:GKO-WZW3} above.

In the $Z$-regular case  we can further 
identify   $\mr{End}(B)\cong \IC[Z]$ as Frobenius algebras.   The group algebra $\IC[Z]$ is endowed with the Frobenius algebra structure corresponding to the convolution product. 
\footnote{Note that $\IC[Z]$ has \underline{two} natural Frobenius algebra structures. See 
the revised version of \cite{Moore:2025tmt} available at \url{https://drive.google.com/file/d/1MFH-pOBjFqHxc7qybWQNU5PQxik0lo4F/view}  for a discussion.   }
%
%
%
%

An important corollary of \eqref{eq:WZW=GKOEndB} is that the partition functions on the torus are related by 
\be\label{eq:RescaleTorus}
Z\left( \mr{WZW}(G,H;k)\right)  = N   Z \left( \mr{GKO}(\fg,\fh;k)\right)  ~,
\ee
where $N=\mr{dim}_{\IC}\,\mr{End}(B)$. Notice that in these cases, $N$ does not depend on the level $k$ unlike the dimension of the Verlinde algebra $V_{\tilde{k}}(H)$ in the $\mr{WZW}(H,H;\tilde{k})$ case. The latter  increases with $\tilde{k}$. 
Specializing to $G= \mr{Spin}(4n)$ and suitable $H$ we recover the vacuum degeneracy observed in  
\cite{Bardakci:1987ee,Altschuler:1987zb}. 

 It is natural to ask whether the result \eqref{eq:WZW=GKOEndB} can be generalized to
 all compact groups.  As we have said, in general  the field identification and fixed-point resolution questions are difficult although, at least in principle, a complete answer to these questions is provided by the results of \cite{Frohlich:2003hg,Frohlich:2003hm}. One of the 
advantages of stating our result as in \eqref{eq:WZW=GKOEndB} is that it suggests a natural generalization, namely for \underline{all} compact Lie groups $G$ and all compact Lie subgroups $H$ we have 
\begin{eqsp}\label{eq:WZW=GKOEndB-gen}
    \mr{WZW}(G,H;k)\cong \mr{GKO}(G,H;k) \, \widetilde{\otimes} \,  \mr{End}(B)~,
\end{eqsp}
where $B$ is defined by the results of \cite{Frohlich:2003hm,Frohlich:2003hg}. 
We have successfully tested this conjecture in some interesting cases. 

One particularly interesting set of cases in which to test \eqref{eq:WZW=GKOEndB-gen}
is that of general \emph{conformal embeddings} \cite{Schellekens:1986mb}. These are the cases where the \tsf{GKO} 
central charge \eqref{eq:GKO-CentralCharge} vanishes. 
In the case of a conformal embedding  $\mr{GKO}(G,H;k)$ is the trivial theory. Thus, 
in this case \eqref{eq:WZW=GKOEndB-gen} identifies the 2d \tsf{TFT}   $\mr{WZW}(G,H;k)$ 
as that defined by    $\mr{End}(B)$ (equipped with a suitable commutative Frobenius algebra structure). 
Another interesting set of examples are the ``Maverick cosets'' \cite{Dunbar:1993hr,dunbar1993characters}.  We have tested 
\eqref{eq:WZW=GKOEndB-gen} in one such example and found that it holds.

This paper leaves open several interesting problems and avenues for further research. 
As we have indicated, it is important to understand better the nature of the ``intertwining'' of the \tsf{GKO} theory with $\mr{End}(B)$ in  \eqref{eq:WZW=GKOEndB}. Moreover, the conjecture
\eqref{eq:WZW=GKOEndB-gen} should be confirmed or disproved. Another interesting set of questions 
concerns whether the result can be derived using the the path integral evaluation of correlation functions of  $\mr{WZW}(G,H;k)$   along the lines of  \cite{Gawedzki:1988nj,Witten:1991mm}. It is important to note that the arguments in these papers are not precise enough to pin down the overall degeneracy factor. 

Our result cries out for an interpretation in terms of 
the quiche picture of topological symmetry in \tsf{QFT}. See \cite{Freed:2022qnc,Moore:2025tmt}
and many references therein for the quiche and related constructions. We note that if we regard $B$ as defining a topological boundary theory in a three-dimensional formulation of the nonchiral theory $\tsf{WZW}(G,H;k)$ then, since $\mr{End}(B)$ has the interpretation of local topological operators there would be a natural interpretation of the topological degeneracy. Moreover, our result could be closely related to the subject of \emph{decomposition} of 2d field theories discussed in the papers \cite{Sharpe:2022ene,Pantev:2005rh,Hellerman:2006zs,Pantev:2005zs}. It would be good to clarify the relation of our results to those papers.

Finally, the extension to \tsf{WZW} models based on noncompact groups is of considerable interest because of the potential applications to string theory and string cosmology. Some brief remarks on such applications can be found in section \ref{sec:implications} below. We also make some brief remarks on the generalized symmetries of gauged \tsf{WZW} models in section \ref{sec:2dQCD}.



The outline of the paper is as follows: in section \ref{sec:GKO_review}, we review the \tsf{GKO} construction in the language of vertex operator algebras. We also describe the construction of category of modules of the \tsf{GKO} \tsf{VOA} using results from category theory. In section \ref{sec:WZW_review}, we review \tsf{WZW} models and their gaugings. In section \ref{sec:IR_G/HWZW}, we discuss the computation of the Hilbert space of gauged \tsf{WZW} model on the circle. We show that \eqref{eq:WZW=GKOEndB} holds at the level of Hilbert space on the circle. We discuss some examples, such as the unitary minimal models, $G/G$-\tsf{WZW} and conformal embeddings, elucidating \eqref{eq:WZW=GKOEndB}. In section \ref{sec:equivariant_H/H}, we interpret the Hilbert space of gauged \tsf{WZW} model as a coupling of \tsf{GKO} Hilbert space equivariant $H/H$-\tsf{WZW} model.   In section \ref{sec:other_examples}, we show that \eqref{eq:WZW=GKOEndB} holds also for other examples that are not $Z$-regular, including parafermions and the case of conformal embeddings.   In section \ref{sec:implications}, we discuss some implications of our results in the cases that coset \tsf{CFT}s are used as building blocks of string theory. We point out some results, conjectures and speculations when the cosets describe black holes and cosmological backgrounds. Finally, in section \ref{sec:2dQCD}, we briefly discuss generalized symmetries in gauged \tsf{WZW} models and its application to 2d Yang-Mills with nonanomalous matter. 


\section*{Acknowledgments}
We would like to thank D. Altschuler, T. Dumitrescu, D. Freed, S. Mukhi, Z. Komargodski, C. Schweigert, N. Seiberg, S. Seifnashri, L. Tizzano   for discussions. 
The work of  G.M. and R.K.S. is supported by the US Department of Energy under grant DE-SC0010008.  E.R. acknowledges
partial support from Israel’s Council for Higher Education grant. We acknowledge the use of LLMs ChatGPT and Claude for useful editorial comments on an earlier version of the paper.
$\,$\\$\,$\\
\centerline{\emph{No part of this paper was written by AI.}}


\section{Review Of The \tsf{GKO} Coset Construction}\label{sec:GKO_review}

In this section, we review the \tsf{GKO} coset construction in a language that will serve us well in the rest of the paper.
\par
Let $(V,Y,\omega^V,\mathbf{1})$ be a \tsf{VOA}. Here $V$ is the vector space of states represented by holomorphic vertex operators, $Y$ is the state to operator map, and $\omega^V$   denotes the ``conformal vector'' - or, in physical terminology, the chiral energy-momentum tensor. Indeed the mode expansion of the vertex 
operator corresponding to this vector  is: 
%
%
\begin{equation}
Y(\omega^V,z)=\sum_{n\in\IZ}L^V(n)z^{-n-2}~,
\end{equation} 
with $L^V(n)$ satisfying the Virasoro algebra  with central charge $c^V$. Finally 
$\mathbf{1}$ is the state mapping under $Y$ to the identity operator on $V$. 

Let   $(U\subseteq V,Y,\omega^U,\mathbf{1})$ be a sub-\tsf{VOA} of $(V,Y,\omega^V,\mathbf{1})$ with
\begin{equation}\label{eq:ModeExpansion}
Y(\omega^U,z)=\sum_{n\in\IZ}L^U(n)z^{-n-2}~,
\end{equation} 
and the operators $L^U(n)$ act on all of   $V$. 
Define the \textit{commutant} of $U$ in $V$ by \cite{10.1215/S0012-7094-92-06604-X,lepowsky2012introduction}
\begin{equation}\label{eq:def_commutant}
\begin{split}
C_V(U)&:=\{v\in V:[Y(v,z_1),Y(u,z_2)]=0~~\text{for all}~~u\in U,z_1,z_2\in\IC\setminus\{0\}\}
\\&=\{v\in V:v_n\cdot u=0~~\text{for all}~~u\in U,n\geq 0\}
\\&=\{v\in V:u_n\cdot v=0~~\text{for all}~~u\in U,n\geq 0\}~,
\end{split}
\end{equation}
where $u_n$ and $v_n$ are operators appearing in the mode expansion of $Y(u,z)$ and $Y(v,z)$ respectively, and $v_n\cdot u$ defines the action of $v_n$ on $u$. The equalities in \eqref{eq:def_commutant} follows from the axioms of a \tsf{VOA}, see \cite{lepowsky2012introduction} for proof. One can also show that \cite{10.1215/S0012-7094-92-06604-X,lepowsky2012introduction}
\begin{eqsp}
    C_V(U)=\mr{Ker}_VL^U(-1)~,
\end{eqsp}
in particular $\textbf{1}\in C_V(U)$. Assuming that $L^V(1)\omega^U=0$, one can show that \cite{10.1215/S0012-7094-92-06604-X,lepowsky2012introduction}
\begin{equation}
L^V(n)=L^U(n),\quad n\geq -1~ \text{on}~~U~. 
\end{equation}
Set 
\begin{equation}
\omega^{V/U}:=\omega^V-\omega^U~.
\end{equation} 
Moreover,
\begin{eqsp}
    C_V(U)\equiv (C_V(U),Y,\textbf{1},\omega^{V/U})
\end{eqsp}
is a sub-\tsf{VOA} of $V$ with central charge 
\begin{eqsp}
    c^{V/U}:=c^V-c^U~.
\end{eqsp}
Also $U\otimes C_V(U)$ is a sub-\tsf{VOA} of $V$ with   conformal vector $\omega^V$.

Two sets of special cases are of particular interest to us: 
First, we  say that a sub-\tsf{VOA} $U$ is \textit{conformally embedded} in $V$ if 
\begin{eqsp}
    c^{V/U}=c^V-c^U=0~.
\end{eqsp}
In this case, if $U,V$ are unitary, selfdual,
\footnote{A \tsf{VOA} $V$ is called selfdual if it is isomorphic to its dual as $V$-modules.}
simple \tsf{VOA}s, and $U$ is conformally embedded in $V$, then one can show that 
\begin{eqsp}
    C_V(U)\cong \IC\textbf{1}~,
\end{eqsp}
a proof will appear in \cite{MRS}.
A second set of special cases is provided when we have a compact connected Lie   subgroup $H< G$ of a compact, connected Lie group $G$. Then we can apply this construction to the affine \tsf{VOA} $V(G;k)$ and its sub-\tsf{VOA} $V(H;\tilde k)$. This defines the \tsf{GKO} coset \tsf{VOA}:
\begin{eqsp}
    V(G,H;k):=C_{V(G;k)}(V(H;\tilde k))~.
\end{eqsp}
When $G,H$ is simply connected, then these \tsf{VOA}s depend only on the Lie algebras. We denote the \tsf{VOA}s in this case by $L_{\fg}(\bk,0):=V(\fg;\bk),L_{\fh}(\wt\bk,0):=V(\fh;\wt\bk)$ and $V(\fg,\fh;\bk)$. 
Recall that for a semisimple Lie algebra $\fg$ the level, as an \tsf{Ad}-invariant form, is quantized 
for each simple factor so we may regard  $\bk\in\IZ_{>0}^{n}$, where $n$ is the number of simple summands in $\fg$. Then the  \tsf{VOA} $L_{\fg}(\bk,0)$ is regular \cite{dong1995regularity} and the finitely many isomorphism classes of irreducible $L_{\fg}(\bk,0)$-modules are in 1-1 correspondence with the \textit{dominant level $\bk$-integrable weights} of $\fg$. We denote this set of dominant integrable weights  by $P^+_{\bk}(\fg)$. See \cite{DiFrancesco:1997nk} for the precise definition. We denote the module corresponding to $\blm\in P^+_{\bk}(\fg)$ by $L_{\fg}(\bk,\hblm)$, where $\hblm$ is the affine weight corresponding to $\blm$. As an $L_{\fh}(\wt{\bk},0)$-module, we have the decomposition
\begin{equation}\label{eq:mod_dec_affine_voa}
L_{\fg}(\bk,\hblm)\cong \bigoplus_{\bm{\mu}\in P^+_{\wt{\bk}}(\fh)}L_{\fg/\fh}(\hblm,\hbmu)\otimes L_{\fh}(\wt{\bk},\hbmu)~,
\end{equation} 
where 
\begin{equation}\label{eq:coset_modules}
L_{\fg/\fh}(\hblm,\hbmu) := \mr{Hom}_{L_{\fh}(\wt{\bk},0)}(L_{\fh}(\wt{\bk},\hbmu),L_{\fg}(\bk,\hblm))~.
\end{equation} 
Moreover, $L_{\fg/\fh}(\hblm,\hbmu)$ is a $V(\fg,\fh;\bk)$-module and 
$L_{\fg/\fh}(0,0)\cong V(\fg,\fh;\bk)$.
  
As discussed in the Introduction, for some pairs $(\hblm,\hbmu)$ 
the module $L_{\fg/\fh}(\hblm,\hbmu)$ can vanish. Moreover it can happen that 
\begin{equation}\label{eq:FieldIdent}
L_{\fg/\fh}(\hblm,\hbmu)\cong L_{\fg/\fh}(\hblm',\hbmu')~,
\end{equation}
for distinct pairs $(\hblm,\hbmu)$ and $(\hblm',\hbmu')$. 
Conditions on $(\hblm,\hbmu)$ under which  $L_{\fg/\fh}(\hblm,\hbmu)$ are nonzero are known as 
\textit{selection rules}. Isomorphisms \eqref{eq:FieldIdent} for distinct pairs are known 
as \textit{field identifications}. (The idea behind the terminology is that there are holomorphic chiral vertex operators $\phi_{(\hblm,\hbmu)}$ and $\phi_{(\hblm',\hbmu')}$ but they should be identified.) Finally, while it often happens that $L_{\fg/\fh}(\hblm,\hbmu)$ are irreducible modules for $V(\fg,\fh;\bk)$, nevertheless it can happen  
that $L_{\fg/\fh}(\hblm,\hbmu)$ is reducible. A typical situation in which this arises is when 
there are fixed points of a simple current action. We next recall the notion of simple currents 
and their action. 

A \emph{simple current}  for a rational \tsf{VOA} $V$ is an irreducible module $J$ for which the \textit{quantum dimension}, defined as 
\begin{eqsp}
    d_J:=\frac{S_{VJ}}{S_{VV}}~,
\end{eqsp}
is 1. Here $S_{W_1W_2}$ is the modular $S$-matrix. Under the fusion product the simple currents form a group.  See \cite{Schellekens:1989uf} for a more detailed discussion of   simple currents.
We  now consider the set
$G_{\mr{id}}$ of pairs of simple currents $(J,J')$ of $L_{\fg}(\bk,0)$ and $L_{\fh}(\wt{\bk},0)$ such that $h_J-h_{J'}\in\IZ$ and the multiplicity space\footnote{By abuse of notation, we are denoting the irreducible module $J$ and the dominant integrable weight corresponding to it by the same symbol $J$.} $L_{\fg/\fh}(J,J')\neq 0$. It turns out that $G_{\mr{id}}$ forms a group and is isomorphic to the common center $Z=Z(G)\cap Z(H)$  if $G,H\neq \mr{E}_8$ at level 2 \cite{MRS}. One description of the 
selection rule given in    \cite{Schellekens:1990xy,Fuchs:1996rq} is the following: First introduce 
\begin{eqsp}
 Q_{J,J'}(\hblm,\hbmu):=Q_{J}(\hblm)-Q_{J'}(\hbmu)~,   
\end{eqsp}
where the \textit{monodromy} $Q_{J}(\hblm)$ is defined by 
\begin{eqsp}
Q_{J}(\hblm):=\left(h_J+h_{\hblm}-h_{J\boxtimes L_{\fg}(\bk,\hblm)}\right)\bmod~1~,
\end{eqsp}
and $h_{\hblm}$ is the conformal dimension of the irreducible $L_{\fg}(\bk,0)$-module $L_{\fg}(\bk,\hblm)$ and $J\boxtimes L_{\fg}(\bk,\hblm)$ is the \textit{fusion product} of modules and $Q_{J'}(\hbmu)$ is defined similarly. It can be proven that, for fixed 
$(\hblm,\hbmu) \in P^+_{\bk}(\fg) \times P^+_{\widetilde \bk}(\fh)$ the map  $\chi_{(\hblm,\hbmu) }: G_{\mr{id}} \to \mr{U(1)}$ defined by $(J,J') \mapsto \exp[2\pi\im  Q_{J,J'}(\hblm,\hbmu)]$ is a character on $G_{\mr{id}}$. Then the selection rule states that 
\be \label{eq:sel_rules_JJ'}
\chi_{(\hblm,\hbmu) }\not= 1\implies   L_{\fg/\fh}(\hblm,\hbmu) = 0 ~ . 
\ee

Field identifications are given by 
\begin{eqsp}\label{eq:field_id_sim_cur}
    L_{\fg/\fh}(\hblm,\hbmu)\cong L_{\fg/\fh}(J\hblm,J'\hbmu)~,\quad (J,J')\in G_{\mr{id}}~,
\end{eqsp}
where $J\hblm$ denotes the dominant integrable weight corresponding to the irreducible
\footnote{The irreducibility of $J\boxtimes L_{\fg}(\bk,\hblm)$ follows from the irreducibility of $L_{\fg}(\bk,\hblm)$ and the fact that  $J$ is a simple current.}
$L_{\fg}(\bk,0)$-module $J\boxtimes L_{\fg}(\bk,\hblm)$ and $J'\hbmu$ is defined similarly. 

An alternative description of selection rules and field identifications follows from
 Chern-Simons theory, as described in \cite{Moore:1989yh}.  Let $W(k,\blm;\gamma)$ be the Wilson line in the Chern-Simons theory $\mr{CS}(G,k)$ along the loop $\gamma$ and $W(-\tilde k, \bmu;\gamma)$ is 
 similarly the line for 
$\mr{CS}(H,-\tilde k)$. The selection rule is simply that the $Z$-action on the two 
Wilson lines must be the same. Put differently,  the product 
$W(k,\blm;\gamma)W(-\tilde k, \bmu;\gamma)$ should be gauge invariant not only under smooth $G\times H$ gauge transformations but also under gauge transformations    that only close up to an element $z\in Z$ around the loop $\gamma$, where $Z$ is embedded diagonally in the center $Z(G) \times Z(H)$. 
On the other hand, field identifications arise from making such a singular gauge transformation 
that only closes up to an element $z\in Z$  around a small loop $\gamma'$ that links $\gamma$.

\par We note parenthetically that field identifications can also be written using  outer automorphism groups of $\hfg,\hfh$ \cite{gepner1989field,DiFrancesco:1997nk}. Every pair of simple current $(J,J')\in G_{\mr{id}}$ corresponds to a pair\footnote{There is a single exception to this correspondence, namely $\hfg=(\hat{\fe}_8)_2$ \cite{Fuchs:1990wb}.} $(\bA,\tilde \bA)\in\mr{Out}(\hfg)\times \mr{Out}(\hfh)$. One can show that for such pairs $(\bA,\tilde \bA)$, 
\begin{eqsp}
     L_{\fg/\fh}(\hblm,\hbmu)\cong L_{\fg/\fh}(\bA\hblm,\tilde\bA\hbmu)~.
\end{eqsp}

Let us illustrate selection rules and field identifications   in an example producing   the 
unitary minimal models 
 \cite{Belavin:1984vu,Friedan:1983xq}. Consider $\wh{\fg}=\wh{\fs\fu}(2)_k\oplus\wh{\fs\fu}(2)_1,k\geq 1$ and $\wh{\fh}=\wh{\fs\fu}(2)_{k+1}$ embedded diagonally in $\fg$. The \tsf{GKO VOA} in this case is the unitary minimal model $\mr{Vir}(c_{k+3,k+2},0)$ with
\footnote{We are denoting the Virasoro Verma module with central charge $c$ and conformal weight $h$ by $\mr{Vir}(c,h)$.} 
central charge $c_{p,q}=1-\frac{6(p-q)^2}{pq}$.  Let us denote the dominant integral weights of $\wh{\fs\fu}(2)_k$ by $i=0,\dots,k$, where $i$ is twice the spin. The simple currents of affine $\fg$ theory are 
\begin{equation}
L_{\fs\fu(2)\oplus\fs\fu(2)}((k,1),(0,0)) \qquad {\rm and} \qquad L_{\fs\fu(2)\oplus\fs\fu(2)}((k,1),(k,1))
\end{equation}
while that of the affine $\fh$ theory are $L_{\fs\fu(2)}(k+1,0),L_{\fs\fu(2)}(k+1,k+1)$. 
The essential observation is that for the affine \tsf{VOA} $L_{\fs\fu(2)}(k,0)$, we have the fusion $k/2\times j=k/2-j$
where $j$ is the spin.
In particular, $k/2\times k/2=0$.
We find that $G_{\mr{id}}=\{\pm\1\}$ with 
\begin{eqsp}
\mathds{1}&:=\left(L_{\fs\fu(2)\oplus\fs\fu(2)}((k,1),(0,0)),L_{\fs\fu(2)}(k+1,0)\right),\\-\1&:=\left(L_{\fs\fu(2)\oplus\fs\fu(2)}((k,1),(k,1)),L_{\fs\fu(2)}(k+1,k+1)\right)~.
\end{eqsp}
The monodromy charge can   be calculated, and we find 
\begin{eqsp}
    Q_{-\1}((\ell_1,\ell_2),\ell_3)=\frac{\ell_1+\ell_2-\ell_3}{2}\bmod 1~.
\end{eqsp}
Thus, the selection rule \eqref{eq:sel_rules_JJ'} implies that 
\begin{eqsp}
    L_{\frac{\fs\fu(2)\oplus\fs\fu(2)}{\fs\fu(2)}}((\ell_1,\ell_2),\ell_3)\neq 0\iff \ell_1+\ell_2-\ell_3\equiv 0\bmod 2~,
\end{eqsp}
which agrees with the explicit computations of \cite{Goddard:1984vk}. Field identifications give 
\begin{eqsp}
L_{\frac{\fs\fu(2)\oplus\fs\fu(2)}{\fs\fu(2)}}((\ell_1,\ell_2),\ell_3)\cong L_{\frac{\fs\fu(2)\oplus\fs\fu(2)}{\fs\fu(2)}}((k-\ell_1,1-\ell_2),k+1-\ell_3)~.    
\end{eqsp}
\par
It is worth noting that there are cases when there are more selection rules and field identifications in addition to \eqref{eq:sel_rules_JJ'} and \eqref{eq:field_id_sim_cur} respectively. 
Such cases are called \textit{Maverick cosets} and examples
were found in \cite{dunbar1993characters,Dunbar:1993hr}.
One Maverick example is discussed in section \ref{sec:maverick}. For conformal embeddings, there are more  selection rules as well as field identifications than what is obtained from $G_{\mr{id}}$.

\par Computational evidence suggests that in the case where all selection rules and field identifications come from 
 $G_{\mr{id}}$,  $L_{\fg/\fh}(\hblm,\hbmu)$ is an irreducible $V(\fg,\fh;\bk)$-module if $(\hblm,\hbmu)$ is not fixed by any element of $G_{\mr{id}}$. 
Otherwise, one needs to decompose $L_{\fg/\fh}(\hblm,\hbmu)$ in terms of irreducible $V(\fg,\fh;\bk)$-modules. This is called \textit{fixed point resolution}, see \cite{Fuchs:1995tq,Schellekens:1989uf} for some progress in this direction. Let us define
\begin{equation}\label{eq:CE_def}
\CE=\left\{(\hblm, \hbmu) \in P^+_{\bk}(\fg)\times P^+_{\wt{\bk}}(\fh): L_{\fg / \fh}(\hblm,\hbmu) \neq 0\right\}~,
\end{equation}
and denote the set of equivalence classes of the $G_{\mr{id}}$-action on $\CE$ by $\CE/\sim$.
Assuming that the set of isomorphism classes of irreducible $V(\fg,\fh;\bk)$-modules are in 1-1 correspondence with $G_{\mr{id}}$-orbits of $\CE$ and there are no fixed points of the $G_{\mr{id}}$-action, we can write the Hilbert space of the \tsf{GKO} theory as 
\begin{eqsp}
    \mathscr{H}_{\fg/\fh}^{\mr{GKO}}:=\bigoplus_{(\hblm,\hbmu)\in\CE/\sim}L_{\fg/\fh}(\hblm,\hbmu)\otimes\wt{L_{\fg/\fh}(\hblm,\hbmu)}~,
\end{eqsp}
where $\wt{L_{\fg/\fh}(\hblm,\hbmu)}$ is the dual space to  $L_{\fg/\fh}(\hblm,\hbmu)$.
It represents the right-moving degrees of freedom. A unitary structure on $L_{\fg/\fh}(\hblm,\hbmu)$ defines an anti-linear isomorphism to the complex conjugate space. For conformal embeddings ($\fh=\fg$ being a special case), the \tsf{GKO} \tsf{CFT} is one dimensional:
\begin{eqsp}
\mathscr{H}_{\fg/\fh}^{\mr{GKO}}=\IC~.    
\end{eqsp}
In the following section, we discuss the \tsf{GKO} coset construction in terms of modular tensor categories associated with \tsf{VOA}s. This serves two purposes: First it is essential when  field identifications and selection rules are not all given by simple currents. Second,  we introduce the algebra object that appears in the main result   \eqref{eq:WZW=GKOEndB} relating the gauged \tsf{WZW} model and \tsf{GKO} coset theory. The reader who would be content just seeing a concrete example might wish to skip to section \ref{sec:IR_G/HWZW}.

\subsection{Categorical Coset Construction}

The issue of selection rules and field identifications can be solved in categorical language. More precisely, using the results of \cite{Frohlich:2003hm} and \cite{kirillov2002q,huang2015braided}, one can construct the category $\CC(\fg,\fh;\bk)$ of modules of the coset \tsf{VOA} $V(\fg,\fh;\bk)$ in terms of the module categories $\CC(\fg;\bk)$ and $\CC(\fh;\wt\bk)$ of $L_{\fg}(\bk,0)$- and $L_{\fh}(\wt\bk,0)$-modules respectively. 

To begin, we have already noted that $V(\fg,\fh;\bk)\otimes L_{\fh}(\wt\bk,0)$ is a sub-\tsf{VOA} of $L_{\fg}(\bk,0)$, so that $L_{\fg}(\bk,0)$ is an \textit{extension} of $V(\fg,\fh;\bk)\otimes L_{\fh}(\wt\bk,0)$. Then, by the results of \cite{kirillov2002q,huang2015braided}, $L_{\fg}(\bk,0)$ is an \textit{  algebra object} in the Deligne tensor product  $\CC(\fg,\fh;\bk)\boxtimes_{\mr{D}}\CC(\fh;\wt\bk)$ and we have an equivalence of braided tensor categories: 
\begin{eqsp}\label{eq:Lg_decom_g/hh}
    \CC(\fg;\bk)\cong \left(\CC(\fg,\fh;\bk)\boxtimes_{\mr{D}}\CC(\fh;\wt{\bk})\right)_{L_{\fg}(\bk,0)}^0~,
\end{eqsp}
where for a tensor category $\CC$ and algebra object $A$, the tensor category of \textit{local} $A$-modules in $\CC$ is denoted $\CC^0_A$.
%
%
%
%
\footnote{One can show that $L_{\fg}(\bk,0)$ is a \emph{haploid} algebra object, meaning that the 
morphism space from $\mathbf{1}$ to $L_{\fg}(\bk,0)$ is one-dimensional.}
One can show moreover that $L_{\fg}(\bk,0)$ is a commutative special symmetric Frobenius algebra object
\footnote{An intuitive physical interpretation of what a ``special symmetric Frobenius algebra object'' means in terms of surface defects is described in \cite{Kapustin:2010if}.} 
in $\CC(\fg,\fh;\bk)\boxtimes_{\mr{D}}\CC(\fh;\wt{\bk})$.

%
%

We would like to give a kind of inverse to \eqref{eq:Lg_decom_g/hh}, constructing 
$\CC(\fg,\fh;\bk) $ out of the \tsf{MTC}s for $\fg$ and $\fh$. Such an inverse follows from 
the   results of \cite{Frohlich:2003hm,Frohlich:2003hg}. 
Let $\CE_0$ be the set
\begin{eqsp}\label{eq:nlm_def}
\CE_0:=\left\{(\hblm,\hbmu)\in\CE: n_{(\hblm,\hbmu)}:=\mr{dim\,Hom}_{\CC(\fg,\fh;\bk)\boxtimes_{\mr{D}}\CC(\fh;\wt{\bk})}\left(V(\fg,\fh;\bk)\boxtimes_{\mr{D}} L_{\fh}(\wt{\bk},\hbmu),L_{\fg}(\bk,\hblm)\right)\neq 0\right\}~,    
\end{eqsp} 
where $L_{\fg}(\bk,\hblm)$ is viewed as an object in the category $\CC(\fg,\fh;\bk)\boxtimes_{\mr{D}}\CC(\fh;\wt{\bk})$ via the decomposition \eqref{eq:mod_dec_affine_voa}. In plain English, $\CE_0$ is the set of pairs $(\hblm,\hbmu)$ such that the multiplicity space $L_{\fg/\fh}(\hblm,\hbmu)$ is contains $n_{(\hblm,\hbmu)}$ copies of $ V(\fg,\fh;\bk)$. 
The object 
\begin{eqsp}\label{eq:B_def_affine}
    B=\bigoplus_{(\hblm,\hbmu)\in \CE_0}n_{(\hblm,\hbmu)}\left(L_{\fg}(\bk,\hblm)\boxtimes_{\mr{D}} L^{\mr{opp}}_{\fh}(\wt{\bk},\hbmu)\right)~,
\end{eqsp}
is a commutative symmetric special Frobenius algebra object in $\CC(\fg;\bk)\boxtimes_{\mr{D}}\CC^{\mr{opp}}(\fh;\wt{\bk})$ and
\begin{eqsp}\label{eq:comm_cat_affine}
    \CC(\fg,\fh;\bk)\cong \left(\CC(\fg;\bk)\boxtimes_{\mr{D}}\CC^{\mr{opp}}(\fh;\wt{\bk})\right)_B^0~,
\end{eqsp}
where $L^{\mr{opp}}_{\fh}(\wt{\bk},\hbmu)$ denotes the object corresponding to $L_{\fh}(\wt{\bk},\hbmu)$ in the opposite category $\CC^{\mr{opp}}(\fh;\wt{\bk})$. 
\par
One can interpret \eqref{eq:comm_cat_affine} in terms of 3d \tsf{TFT} as follows. There is a construction of a 3d \tsf{TFT} from an \tsf{MTC} \cite{witten1989quantum,Reshetikhin:1991tc,turaev1992modular,turaev1994quantum}.
 By the results of \cite{Kapustin:2010if,Freed:2020qfy}, 
 any commutative symmetric special Frobenius algebra object $A$ in an \tsf{MTC} $\CM$ can be interpreted as a topological boundary condition in the corresponding 3d \tsf{TFT}. Because of the relation to condensed matter physics \cite{Moore:1991ks,Kitaev:2005hzj},
 the objects in an \tsf{MTC} are referred to as \emph{anyons}. Restricting to the category $\CM_A^0$ of local $A$-modules in $\CM$ is then interpreted as restricting to bulk anyons which do not ``see'' the topological boundary theory. Another way one can interpret $\CM_A^0$ is to make the bulk anyon corresponding to $A$ invisible. The existence of an $A$-module structure means that the $A$-anyon can fuse with $A' \in \CM_A^0$ to produce $A'$ - and in this sense $A$ is ``invisible''.  This is often called ``anyon condensation'' in literature \cite{bais2009condensate,kong2014anyon}. 

\par Note that the 3d \tsf{TFT} corresponding to the \tsf{MTC} $\CC(\fg;\bk)\boxtimes\CC^{\mr{opp}}(\wt\bk;\fh)$ is the $(G_{\bk}\times H_{-\wt{\bk}})$-Chern-Simons theory.
In many cases, the above relation is closely related to   gauging the common center of $G$ and $H$, whose relevance was noted in \cite{Moore:1989yh}. More precisely, the chiral \tsf{GKO} theory can be obtained as a boundary theory of the Chern-Simons theory with gauge group $(G_{\bk}\times H_{-\wt{\bk}})/Z$, where $Z:=Z(G)\cap Z(H)$ is the common center of $G$ and $H$. There is a map 
$Z \to P^+_{\bk}(\fg) \times P_{\wt\bk}^+(\fh)$ and we can view $B$ as a sum over the   anyons in the $G_{\bk}\times H_{-\wt{\bk}}$ Chern-Simons theory corresponding to the image of this map. 

In the case that the \tsf{GKO} theory is $Z$-regular we can simplify the formula \eqref{eq:B_def_affine} for $B$ to 
\be\label{eq:SimpleB}
B = \bigoplus_{(J,J')\in G_{\mr{id}}}  (J \boxtimes_{\mr{D}} (J')^{\mr{opp}})~. 
\ee
where we use $J,J'$ to denote the corresponding modules of the vertex operator algebras. 

%
%
%

In order to set up our main calculation, we first review the  \tsf{WZW} models and their gauging.

\section{Gauged \tsf{WZW} Model}\label{sec:WZW_review}

In this section, we review \tsf{WZW} models and their gauging which will serve the purpose of introducing some notations required for the ultimate goal of computing the Hilbert space of the gauged \tsf{WZW} model on the circle. \par Let $G$ be a compact, semisimple, connected and simply connected Lie group. We can write 
\begin{eqsp}
    G\cong\prod_{i=1}^nG_i~,
\end{eqsp}
with each $G_i$ compact,  connected,  simply connected and simple. Since $H^4(BG_i,\IZ)\cong \IZ$, the level $\bk\in H^4(BG,\IZ)$ can be taken to be a tuple $\bm{k}=(k_1,\dots,k_n)\in(\IZ_{>0})^n$. Let $\langle\cdot,\cdot\rangle_{\bm{k}}$ be the Killing form on $\fg:=\mr{Lie}(G)$
extended to the complexification $\fg^{\IC}:=\oplus_i\fg_i^{\IC}$ normalized such that half the length-squared of long coroots of $\fg_i$ is equal to $k_i$.
For any Riemann surface $\Sigma$ and a smooth map $g:\Sigma\longrightarrow G$, define the functional
\begin{equation}\label{eq:WZW_action}
    S_{G,\bk}(g):=\frac{\im}{4\pi}\int_\Sigma\rd^2z~ \langle g^{-1}\partial_z g,g^{-1}\partial_{\bar{z}}g\rangle_{\bm{k}}-\im\Gamma_{\mr{WZ}}(\Sigma;g)~,
\end{equation}
where the usual definition of the Wess-Zumino (\tsf{WZ}) term is written as: 
\begin{equation}\label{eq:WZ_term}
    \Gamma_{\mr{WZ}}(\Sigma;g):=\frac{1}{24\pi}\int_{M_3}\left\langle\wt{g}^{-1}\rd\wt{g},\left[\wt{g}^{-1}\rd\wt{g},\wt{g}^{-1}\rd\wt{g}\right]\right\rangle_{\bk}~,
\end{equation}
where 
is  $M_3$ is a 3-manifold such that $\pd M_3=\Sigma$ and $\wt{g}:M_3\longrightarrow G$ is an extension
 of the map $g$, i.e., $\wt{g}|_{\partial M_3}=g$.
\footnote{\label{foot:gen_WZ_term}When $G$ is simply connected, then the existence of such an extension follows from $\pi_1(G)=\pi_2(G)=0$. For general $G$ the term has been discussed in  \cite{Gawedzki:1987ak,Felder:1988sd,Axelrod:1989xt,Hori:1994nc}. The most incisive definition - and the one that extends to the gauged case - uses differential cohomology. See   \cite{Moore:2025tmt} and references therein for the ungauged case. The extension to the gauged \tsf{WZ}-term is formulated in a way similar to the four-dimensional case studied in   \cite{Freed:2006mx}.}
The action is manifestly conformally invariant. The action has a $(G\times G)$-symmetry which acts on $g$ as 
\begin{equation}\label{eq:WZWGxG_sym}
    (g_L,g_R)\cdot g=g_Lgg_R^{-1}~, \quad (g_L,g_R)\in G\times G~.
\end{equation}
Up to a constant multiple, the currents for these symmetries in complex coordinates are
\begin{equation}\label{eq:currents_WZW}
J= g^{-1}\partial_z g,\quad \wt{J}=-g^{-1} \partial_{\bar{z}} g~.
\end{equation}
Given a subgroup $H<G$, we can gauge the diagonally embedded $H\subset G\times G$. This acts on the \tsf{WZW} field by conjugation. We then should sum over $H$-bundles and integrate over $H$-connections.  
For a general principal $H$-bundle $P\to \Sigma$ the \tsf{WZW} field is a section of the associated 
bundle $P\times_H H$. In the case that $P=\Sigma \times H$ the $H$-gauged \tsf{WZW} action is given by 
\begin{equation}\label{eq:action_gauged_WZW}
\begin{split}
    S_{G,\bk}(g,A):=S_{G,\bk}(g)+\frac{\im}{2\pi}\int_\Sigma \rd^2z&\left(\left\langle A_z,g^{-1}\pd_{\bar{z}} g\right\rangle_{\bk}+\left\langle g\pd_z g^{-1},A_{\ov{z}}\right\rangle_{\bk}\right.
    \\
    &\left.+\left\langle gA_zg^{-1},A_{\ov{z}}\right\rangle_{\bk}-\left\langle A_z,A_{\ov{z}}\right\rangle_{\wt{\bk}}\right)~,
\end{split}    
\end{equation}
where the $H$-gauge field is written as 
\begin{equation}
    A=A^{(1,0)}+A^{(0,1)}~, \quad A^{(1,0)}=-(A^{(0,1)})^*~,
\end{equation}
where $A^{(1,0)},A^{(0,1)}$ is the $\rd z$ and $\rd\ov{z}$ parts:
\begin{eqsp}
    A^{(1,0)}=A_z\rd z~,\quad A^{(0,1)}=A_{\ov{z}}\rd\ov{z}~.
\end{eqsp}

\section{The State Space On The Circle Of The Gauged \tsf{WZW} Models}\label{sec:IR_G/HWZW}

\subsection{General Remarks}\label{subsec:GenRemarksStateSpace}

In this section, we will describe the finite-energy gauge-invariant states in the $G/H$-\tsf{WZW} model when quantized on a cylinder with a spatial circle. We will add the Maxwell term to the $G/H$-\tsf{WZW} action and then take the gauge coupling to infinity to flow to the $G/H$-\tsf{WZW} model as originally suggested in \cite{rabinovici1988aspects}. We will then identify the \textit{finite-energy gauge invariant states} of the theory in the $e^2 \to \infty$ limit. (More properly, if $L$ is the radius of the circle we are taking $e^2 L^2 \to \infty$.) 
We will see that the Ishibashi states of the $H$-\tsf{WZW} model naturally appear when we impose gauge invariance.

For simplicity, we asssume that $G,H$ are compact, connected, semisimple and simply connected.  
Thus, we consider the action
\begin{eqsp}\label{eq:action-G/H-Maxwell}
    S(A,g):=\frac{1}{e^2}\int_{\Sigma}\left\| F(A)\right\|_{\wt{\bk}}^2+S_{G,\bk}(g,A)~,
\end{eqsp}
where $F(A)$ is the curvature of the $H$-connection $A$, and
\begin{eqsp}
    \left\| F(A)\right\|^2_{\wt{\bk}}:=\left\langle F(A), \star F(A))\right\rangle_{\wt{\bk}}~,
\end{eqsp}
and $S_{G,\bk}(g,A)$ is the $G/H$-\tsf{WZW} action given, in the topologically trivial sector, by \eqref{eq:action_gauged_WZW}. (See footnote \ref{foot:gen_WZ_term} for the extension to topologically nontrivial sectors.)  The gauge coupling $e^2$ has dimensions of mass${}^2$ and triggers a \tsf{RG} flow. We want to identify the  gauge invariant states of the theory governed by \eqref{eq:action-G/H-Maxwell} that remain at finite energy in the $e^2\to\infty$ limit.

The Hilbert space of the theory is given by 
%
%
\begin{eqsp}\label{eq:Invt-G/H-HilbSpace}
%
\mathscr{H}^{\mr{WZW}\otimes\mr{YM}}_{G}\cong
\left(\mathscr{H}_{G}^{\mr{WZW}}\otimes L^2(\CA)\right)^{LH}~,
\end{eqsp}
where $\mathscr{H}_{G}^{\mr{WZW}}$ is the Hilbert space of $G$-\tsf{WZW} model,  $L^2(\CA)$ is the space of square integrable functions on the affine space $\CA$ of $H$-connections, and the gauge group is just the loop group $LH$. The superscript on the RHS of \eqref{eq:Invt-G/H-HilbSpace} means that we project onto the gauge invariant states. The Hamiltonian of the theory is given by
\footnote{We are considering the Hamiltonian on the cylinder with units chosen so the circumference of the cylinder is normalized to $2\pi$. By $L_0^{\fg}+\bar{L}_{0}^{\fg}$ we mean the Hamiltonian of the \tsf{WZW} theory in the presence of a background gauge field on the circle. As we will see, finite-energy states in the $e^2\to \infty$ limit must lie in the subspace of wavefunctions that are independent of $A_\sigma$. The compression of  $L_0^{\fg}+\bar{L}_{0}^{\fg}$  to this subspace is the standard \tsf{WZW} Hamiltonian in the absence of a background field.   }  
\begin{eqsp}
    H=(L_0^{\fg}+\bar{L}_{0}^{\fg})+e^2\oint_{S^1} \rd \sigma\,\frac{\delta}{\delta A^a(\sigma)}\frac{\delta}{\delta A^a(\sigma)}~, 
\end{eqsp}
where $a$ runs over an orthonormal basis for $\fh$ with respect to a Killing form. (The normalization of the Killing form does not matter because we will take the $e^2\to \infty$ limit.) 
Consider a state of the form $\psi \otimes f$ where $\psi \in \mathscr{H}_{G}^{\mr{WZW}}$
and $f\in L^2(\CA)$. Finite energy states in the $e^2\to\infty$ limit must satisfy: 
\begin{eqsp}
    \lim_{e^2\to\infty}e^2\oint_{S^1} \rd \sigma\,\frac{\delta}{\delta A^a(\sigma)}\frac{\delta}{\delta A^a(\sigma)}f(A)<\infty~.
\end{eqsp}
This requires
\begin{eqsp}
 \oint_{S^1} \rd \sigma\,\frac{\delta}{\delta A^a(\sigma)}\frac{\delta}{\delta A^a(\sigma)}f(A)=0~.  
\end{eqsp}
This forces $f(A)$ to be a constant as a function of $A$.
We thus conclude that the Hilbert space of gauge invariant states that have finite energy in the $e^2\to \infty$ limit 
is the subspace of \eqref{eq:Invt-G/H-HilbSpace} given by 
\begin{eqsp}
 \left(\mathscr{H}_{G}^{\mr{WZW}}\otimes\IC\right)^{LH}~\cong\left(\mathscr{H}_{G}^{\mr{WZW}}\right)^{LH}~,     
\end{eqsp} 
where $\IC$ denotes the space of constant functions on $\CA$.
We will now determine the gauge invariant states determined by imposing the Gauss law. We parametrize the cylinder by $(\sigma,\tau)$ where $\sigma\sim\sigma+2\pi$ parametrizes the spatial direction and $\tau$ is the Euclidean time. The conformal map to the 
punctured plane is given by $z=e^{-\im\sigma+\tau}$. Note that 
\begin{eqsp}\label{eq:d_sigmad_z}
    \partial_\sigma=\im\left(\bar{z}\partial_{\bar{z}}-z\partial_z\right)~.
\end{eqsp} 
We are gauging the symmetry $H \hookrightarrow G \times G$ embedded diagonally and acts on the \tsf{WZW} field $g \in \mr{Map}(\Sigma, G)$ as
\begin{equation}\label{eq:G/H_sym}
g \mapsto h g h^{-1}~.
\end{equation}
Let us write the gauged \tsf{WZW} action \eqref{eq:action_gauged_WZW} for this diagonal gauging of $H$ in terms of differential forms. For $H$ connected all principal $H$ bundles on the circle or cylinder are topologically trivial so we may write:  
\begin{eqsp}\label{eq:gauged_WZW_forms}
S_{G,\bk}(g,A)=S_{G,\bk}(g)+\frac{\im}{2\pi}\sum_{a'=1}^{\mr{dim}(\fh)}\int_\Sigma A^{a'}\wedge\langle t^{a'},g^{-1}\rd g\rangle_{\bk}
    &+\frac{\im}{4\pi}\sum_{a',b'=1}^{\mr{dim}(\fh)}\int_{\Sigma}(A^{a'}\wedge A^{b'})\langle gt^{a'}g^{-1},t^{b'}\rangle_{\bk}
    \\&+\frac{\im}{4\pi}\sum_{a',b'=1}^{\mr{dim}(\fh)}\int_{\Sigma}(A^{a'}\wedge A^{b'})\left\langle t^{a'},t^{b'}\right\rangle_{\wt{\bk}}~,    
\end{eqsp}
where $\{t^{a'}\}_{a'=1}^{\mr{dim}(\fh)}$ is an orthonormal basis of $\fh$. 
%
%
The Gauss law constraint is imposed on physical states as an operator equation of the field constraint 
\begin{eqsp}
    \frac{\delta S_{G,\bk}(g,A)}{\delta A_\tau}=0~.
\end{eqsp}
Note that this constraint is to be imposed on the states in the Hilbert space at fixed Euclidean time $\tau=\tau_0$. We will take $\tau_0=0$. 
From \eqref{eq:gauged_WZW_forms}, we see that the Gauss law constraint on physical states is 
\begin{eqsp}
    \wh{\CG}|\mr{phys}\rangle=0~,
\end{eqsp}
where 
\begin{eqsp}
\wh{\CG}:=\sum_{a'=1}^{\mr{dim}(\fh)}\left\langle t^{a'},g^{-1}\partial_\sigma g\right\rangle_{\bk}+\frac{1}{2}\sum_{a',b'=1}^{\mr{dim}(\fh)}(A^{b'}_{\sigma}-A^{a'}_{\sigma})\left(\langle gt^{a'}g^{-1},t^{b'}\rangle_{\bk}+\langle t^{a'},t^{b'}\rangle_{\bk}\right) ~.   
\end{eqsp}
The sum over   $a',b'$ vanishes due to the antisymmetry under $a'\leftrightarrow b'$. 
(This follows from  the 
$\mr{Ad}$-invariance   of the symmetric bilinear form $\langle\cdot,\cdot\rangle_{\bk}$.)   
The Gauss law   becomes 
\begin{eqsp}
\sum_{a'=1}^{\mr{dim}(\fh)}\left\langle t^{a'},g^{-1}\partial_\sigma g\right\rangle_{\bk}|\mr{phys}\rangle=0~.    
\end{eqsp}
From \eqref{eq:currents_WZW} and \eqref{eq:d_sigmad_z}, the Gauss law   takes the form 
\begin{eqsp}
   \sum_{a'=1}^{\mr{dim}(\fh)} \left(z\langle t^{a'},J(z)\rangle_{\bk}+\bar{z}\langle t^{a'},\wt{J}(\bar{z})\rangle_{\bk}\right)|\mr{phys}\rangle=0~.
\end{eqsp}
We identify 
\begin{eqsp}
J_{\fh}(z):=\sum_{a'=1}^{\mr{dim}(\fh)} \langle t^{a'},J(z)\rangle_{\bk}~,\quad    \wt{J}_{\fh}(\bar{z}):=\sum_{a'=1}^{\mr{dim}(\fh)}\langle t^{a'},\wt{J}(\bar{z})\rangle_{\bk}~, 
\end{eqsp}
with the Kac-Moody current for $\fh$. Expanding the $\fh$ Kac-Moody currents into modes, the Gauss law takes the form 
\begin{eqsp}
    \sum_{n\in\IZ}e^{-\im n\sigma}(J^{a'}_n+\wt{J}_{-n}^{a'})|\mr{phys}\rangle=0~,\quad a'=1,\dots,\mr{dim}(\fh)~.
\end{eqsp}
Equivalently, the Gauss law   can be written as 
\begin{eqsp}\label{eq:gauge_inv_cond}
    (J^{a'}_n+\wt{J}_{-n}^{a'})|\mr{phys}\rangle=0~,\quad n\in\IZ~,~~a'=1,\dots,\mr{dim}(\fh)~.
\end{eqsp}
Note that \eqref{eq:gauge_inv_cond} imposes gauge invariance only for the infinitesimal gauge transformations. Because $\pi_0(LH) \cong \pi_1(H)$ we will now restrict to the case that 
$H$ is simply connected. In this case, for $H$ semisimple the image of the exponential map in $LH$ is dense \cite[Proposition 3.5.3]{PressleySegal1986}. 
Thus \eqref{eq:gauge_inv_cond}  is also sufficient for invariance under all gauge transformations. 
All in all, we want to compute the following space:
\begin{eqsp}
    \mathscr{H}^{\mr{WZW}}_{G/H}:=\left\{|\psi\rangle\in\mathscr{H}^{\mr{WZW}}_{G}:(J^{a'}_{n}+\wt{J}^{a'}_{-n})|\psi\rangle=0\right\}~.
\end{eqsp}
To proceed, we use the decomposition\footnote{Note that we are using the fact that $G$ is simply connected in the first line of \eqref{eq:HgWZW_decomp}. If $G$ is not simply connected, then the Hilbert space changes as described in \cite{Moore:1989yh}.}:
\begin{eqsp}\label{eq:HgWZW_decomp}
\mathscr{H}_{G}^{\mr{WZW}}&=\bigoplus_{\blm \in P_{\bk}^{+}(\fg)} L_{\fg}(\bk, \hblm)\otimes\wt{L_{\fg}(\bk, \hblm)} \\&\cong \bigoplus_{\substack{(\blm, \bmu) \in \CE\\(\blm, \tilde{\bmu}) \in \CE}} \left(L_{\fg / \fh}(\hblm, \hbmu) \otimes L_{\fh}(\widetilde{\bk}, \hbmu)\right)\otimes \left(\wt{L_{\fg / \fh}(\hblm, \hat{\tilde{\bmu}})} \otimes \wt{L_{\fh}(\widetilde{\bk}, \hat{\tilde{\bmu}})}\right)
\\&\cong \bigoplus_{\substack{(\blm, \bmu) \in \CE\\(\blm, \tilde{\bmu}) \in \CE}} \left(L_{\fg / \fh}(\hblm, \hbmu) \otimes \wt{L_{\fg / \fh}(\hblm, \hat{\tilde{\bmu}})} \right)\otimes \left( L_{\fh}(\widetilde{\bk}, \hbmu)\otimes \wt{L_{\fh}(\widetilde{\bk}, \hat{\tilde{\bmu}})}\right)~.
\end{eqsp}
So, the solutions to 
\begin{eqsp}
    (J^{a'}_n+\wt{J}_{-n}^{a'})|\psi\rangle=0~,\quad n\in\IZ~,~~a'=1,\dots,\mr{dim}(\fh)~,\quad |\psi\rangle\in \mathscr{H}_{G}^{\mr{WZW}}~, 
\end{eqsp}
can be written as 
\begin{eqsp}
  |\psi\rangle=\sum_{\substack{(\blm, \bmu) \in \CE\\(\blm, \tilde{\bmu}) \in \CE}}\left|\psi^{\fg/\fh}_{(\blm,\bmu),(\blm,\tilde{\bmu})}\right\rangle \otimes |\bmu,\tilde{\bmu}\rrangle~, 
\end{eqsp}
where 
\begin{eqsp}
\left|\psi^{\fg/\fh}_{(\blm,\bmu),(\blm,\tilde{\bmu})}\right\rangle \in L_{\fg / \fh}(\hblm, \hbmu) \otimes \wt{L_{\fg / \fh}(\hblm, \hat{\tilde{\bmu}})}~,\quad   |\bmu,\tilde{\bmu}\rrangle\in L_{\fh}(\widetilde{\bk}, \hbmu)\otimes \wt{L_{\fh}(\widetilde{\bk}, \hat{\tilde{\bmu}})}~,
\end{eqsp}
such that $|\bmu,\tilde{\bmu}\rangle$ satisfies the symmetry-preserving constraint 
\begin{eqsp}\label{eq:Ishibashi_non-diag}
\left[(J^{a'}_n\otimes \mathds{1})+(\mathds{1}\otimes \wt{J}^{a'}_{-n})\right]|\bmu,\tilde{\bmu}\rrangle=0~.    
\end{eqsp}
The constraint \eqref{eq:Ishibashi_non-diag} implies that $\tilde{\bmu} = \bmu$ and we note that \eqref{eq:Ishibashi_non-diag} is the Ishibashi constraint studied in context of conformal boundary conditions in \cite{Ishibashi:1988kg}. 
The solution to Ishibashi constraints were given in \cite{Ishibashi:1988kg}: they are in 1-1 correspondence with the irreducible modules of the chiral algebra. We denote the Ishibashi state
\footnote{To be precise, Ishibashi states are not really ``states'' because they are not normalizable. However they become normalizable when acted on by the contraction operator $q^{L_0} \bar q^{\tilde L_0}$ with $\vert q \vert < 1$. The Ishibashi states are in the dual of the space spanned by such contracted states.  We suggest that the formalism of \textit{rigged} Hilbert spaces formulated in \cite{GelfandShilov1964,GelfandVilenkin1964} is a suitable language for discussing these states, but will not pursue that interesting line of enquiry here. By Ishibashi ``state'' we really mean the line through the vector $\vert \bmu \rrangle$. }
corresponding to the irreducible $L_{\fh}(\wt\bk,0)$-module $L_{\fh}(\wt\bk,\hbmu)$ by 
\be
|\bmu\rrangle \in L_{\fh}(\widetilde{\bk}, \hbmu)\otimes \wt{L_{\fh}(\widetilde{\bk}, \hat{\bmu})}~.
\ee
With this understood the   space of gauge-invariant states takes the form
\begin{eqsp}\label{eq:GmodH-WZW-space}
   \mathscr{H}^{\mr{WZW}}_{G/H}&:=\bigoplus_{(\blm,\bmu)\in\CE} \left(L_{\fg / \fh}(\hblm, \hbmu) \otimes \wt{L_{\fg / \fh}(\hblm, \hbmu)}\right)\otimes \IC|\bmu\rrangle
   \\&\cong\bigoplus_{(\blm,\bmu)\in\CE/\sim} \left[\left(L_{\fg / \fh}(\hblm, \hbmu) \otimes \wt{L_{\fg / \fh}(\hblm, \hbmu)}\right)\otimes \bigoplus_{(J,J')\in G_{\mr{id}}}\IC|J'\bmu\rrangle\right]~.
\end{eqsp} 

\bigskip 
\begin{remark}\label{rem:IshibashiGauss}
  Ishibashi states are usually encountered as closed string states dual (via dual channels on the cylinder and the annulus) to symmetry-preserving boundary conditions.  For example, in the case where $H=\mr{U(1)}$, the constraint \eqref{eq:Ishibashi_non-diag} corresponds to  a Dirichlet boundary condition. The fact that we have encountered these boundary states via Gauss' law is novel. Note that we are encountering Ishibashi - not Cardy - states \cite{Cardy:1989ir}. This raises the interesting question of how our result fits with  the incorporation of local boundary conditions - but that is beyond the scope of this work. 
\end{remark}
\bigskip 
\begin{remark}\label{rem:H_non_sc}
When $H$ is not simply connected, then one needs to impose Gauss' law for components of $LH$ not connected to identity. This is quite important for us since many examples that we discuss here fall in this category. The generalization of \eqref{eq:GmodH-WZW-space} to this case will appear in \cite{MRS}.    
\end{remark}
In the following we will interpret equation \eqref{eq:GmodH-WZW-space} in various ways. 
To begin, we consider the $Z$-regular case, which  was defined in the introduction. 
Bearing in mind that $Z\cong G_{\mr{id}}$ it follows from the 
definition given in the introduction  that  there is a 1-1 correspondence between the set   of  isomorphism classes of irreducible $V(\fg,\fh;\bk)$-modules and the set of orbits of the  $G_{\mr{id}}$-action on $\CE$. 

We denote the set of isomorphism classes of irreducible $V(\fg,\fh;\bk)$-modules
by $\CA_{\fg/\fh}$ and a general element of this set is denoted by $r$.  For 
each  $r\in \CA_{\fg/\fh}$ there are projection maps 
\begin{eqsp}
    \pi_1:\CO_r\longrightarrow P^+_{\bk}(\fg)~,\quad \pi_2:\CO_r\longrightarrow P^+_{\wt{\bk}}(\fh)~,
\end{eqsp}
where $\CO_r$ is the $G_{\mr{id}}$-orbit of $\CE$ corresponding to the irrep $r\in \CA_{\fg/\fh}$. We then have 
\begin{eqsp}\label{eq:G/H_space_RepS}
    \mathscr{H}^{\mr{WZW}}_{G/H}\cong \bigoplus_{r\in\CA_{\fg/\fh}} \left[\left(W^r \otimes \wt{W}^r\right)\otimes\bigoplus_{\bmu\in\pi_2(\CO_r)}\C[\CS_{\bmu}]\right]~, 
\end{eqsp}
where $\CS_{\bmu}< G_{\mr{id}}$ is the stabilizer subgroup\footnote{$G_{\mr{id}}$ acts on $P^+_{\wt\bk}(\fh)$ as $(J,J')\cdot\bmu=J'\bmu$.} of $\bmu$ and $\C[\CS_{\bmu}]$ is the group algebra of $\CS_{\bmu}$. Let $\CO_r=[(\hblm,\hbmu)]$ be the $G_{\mr{id}}$-orbit of $(\hblm,\hbmu)\in\CE$. We have the isomorphism
\begin{eqsp}\label{eq:non_nat_iso}
    \bigoplus_{(J,J')\in G_{\mr{id}}}\IC|J'\bmu\rrangle\cong \bigoplus_{\bmu'\in\pi_2(\CO_r)}\C[\CS_{\bmu'}]~,
\end{eqsp}
although it is not canonical unless $\bmu=0$.
Under our assumption, we see that - for each $r$ -  there is an isomorphism of vector spaces: 
\begin{eqsp}
\bigoplus_{\bmu\in\pi_2(\CO_r)}\C[\CS_{\bmu}]\cong\IC[G_{\mr{id}}]~.    
\end{eqsp}
and hence each ``sector'' labeled by $r$ has the same degeneracy - as a vector space. 
Next, in this case, the algebra object $B$ is given by the simple formula \eqref{eq:SimpleB}. 
%
%
Thus, we have 
\begin{eqsp}\label{eq:CGid=EndB}
\IC[G_{\mr{id}}]\cong  \mr{End}(B)\cong\bigoplus_{(J,J')\in G_{\mr{id}}}\IC f_{(J,J')}~.    
\end{eqsp}
where 
\begin{eqsp}\label{eq:fJJ'_def}
f_{(J,J')}:=\iota_{(J,J')}\circ r_{(J,J')}~,    
\end{eqsp}
where 
\begin{eqsp}\label{eq:iJJ'rJJ'_def}
\iota_{(J,J')}&:J\boxtimes_{\mr{D}}(J')^{\mr{opp}}\longrightarrow B~,
    \\
    r_{(J,J')}&:B\longrightarrow J\boxtimes_{\mr{D}}(J')^{\mr{opp}}~,    
\end{eqsp}
are the inclusion and projection maps respectively. Thus the topological degeneracy spaces, 
for each $r$, are   isomorphic to $\mr{End}(B)$ but only as vector spaces. 

We would like to reformulate our result so that we can easily generalize it beyond 
the $Z$-regular case. In the $Z$-regular case the topological degeneracy is 
controlled by the  group algebra $\IC[G_{\mr{id}}]$. This algebra has a natural structure of a commutative Frobenius algebra structure defined by convolution. (It is commutative since 
$G_{\mr{id}}$ is Abelian.) Therefore 
$\IC[G_{\mr{id}}]$ determines a   2d \tsf{TFT} and we can interpret the topological degeneracy as 
arising from this 2d \tsf{TFT}.  Now, if we change our focus and regard the topological degeneracy 
as   $\mr{End}(B)$ then we observe that - 
even without the assumption of $Z$-regularity - it is still true that  $\mr{End}(B)$ admits the structure of a \underline{commutative} Frobenius algebra (see equation \eqref{eq:mult_EndB} below). Moreover, in the $Z$-regular case  equation \eqref{eq:CGid=EndB} is an isomorphism of Frobenius algebras, if we endow $\IC[G_{\mr{id}}]$ with the Frobenius algebra structure corresponding to the convolution product. 
\par
The (Frobenius) algebra structure on $\mr{End}(B)$ is induced from that on $B$. For example, the multiplication is given by 
\begin{eqsp}\label{eq:mult_EndB}
    f\cdot g:=\mu_B\circ (f\boxtimes g)\circ \Delta_B~,
\end{eqsp}
where $\mu_B,\Delta_B$ are multiplication and comultiplication on $B$. The Frobenius pairing is defined by 
\begin{eqsp}\label{eq:Frob_trace_EndB}
    \langle f,g\rangle:=\varepsilon_B\circ( f\cdot g)\circ\eta_B=\varepsilon_B\circ\mu_B\circ (f\boxtimes g)\circ\Delta_B\circ\eta_B~,
\end{eqsp}
where $\eta_B,\varepsilon_B$ are the unit and counit morphisms on $B$. The fact that the multiplication \eqref{eq:mult_EndB} on $\mr{End}(B)$  is \underline{commutative} is extremely surprising and is related to the fact that $B$ is commutative and cocommutative as an algebra object. 
%
%

Thus, we arrive at our main result, as announced in the introduction: 

\begin{thm}\label{thm:main_thm}
Let $H < G$ be a pair of connected, simply-connected, compact Lie groups  such that $V(G,H;k)$ is $Z$-regular, then  
\begin{eqsp}\label{eq:Main-EndB-statement}
\mathscr{H}^{\mr{WZW}}_{G/H}\cong \mathscr{H}^{\mr{GKO}}_{\fg/\fh} \, \widetilde{\otimes}\, \mr{End}(B)~, 
\end{eqsp}
where $B$ is the algebra object defined in \eqref{eq:B_def_affine} and $\mr{End}(B)$ is endowed with the commutative Frobenius algebra structure given in \eqref{eq:mult_EndB}, \eqref{eq:Frob_trace_EndB}.
\end{thm}
\begin{remark}\label{rem:EndB-mod}
Recall that the isomorphism \eqref{eq:non_nat_iso} is not natural for $\bmu\neq 0$. But for $\bmu\neq 0$, the left-hand-side of \eqref{eq:non_nat_iso} carries the structure of a rank-1 free $\C[G_{\mr{id}}]$-module. (This is the analog for algebras of a  torsor for a group.)  The action is given by 
\begin{eqsp}
    (J,J')\cdot |\tilde{J}'\cdot\hbmu\rrangle=|(J'\tilde{J}')\cdot\hbmu\rrangle~.
\end{eqsp}
Thus, an alternative - perhaps more cogent - way of stating the result is that the vacuum sector of $\mr{WZW}(G,H;\bk)$ is the vacuum sector of $\mr{GKO}(\fg,\fh;\bk)$ tensored with the 2d \tsf{TFT}
determined by $\mr{End}(B)$ and other sectors of $\mr{WZW}(G,H;\bk)$ are related to the corresponding sectors of $\mr{GKO}(\fg,\fh;\bk)$ by tensoring with rank-1 free $\mr{End}(B)$-modules.
\end{remark}
We now provide some examples elucidating \eqref{eq:Main-EndB-statement}.

\subsection{Examples}\label{subsec:Examples}
In this section,  we discuss some examples of $Z$-regular gauged \tsf{WZW} models. We show that the topological degeneracy is given by the group algebra of $Z$. We identify the topological degeneracy with $\mr{End}(B)$. Note that we do not need the categorical coset construction results to identify the \tsf{TFT}s in the examples discussed in section \ref{subsec:Examples}. 
\subsubsection{Unitary Minimal Models}
We consider two different coset realizations of the unitary minimal models.

\paragraph{Example 1: $G=\mr{SU}(2)_k\times \mr{SU}(2)_1,H=\mr{SU(2)}_{k+1}$.}
The \tsf{GKO} result for this coset reproduces the central charges of the 
unitary minimal models \cite{Belavin:1984vu,Friedan:1983xq}:   
\be 
c_{k}=1-\frac{6}{(k+2)(k+3)} ~ . 
\ee
%
%
The topological sector is the group algebra $\IC[\IZ_2]$ of the common center, for all $k$. Let us demonstrate this explicitly for $k=1$, which corresponds to the Ising model. 
Let us label the dominant integral weights of $\mr{SU(2)}_k$ by $i=0,\dots,k$. That is, we label them by twice the spin of the lowest weight vector. Let us denote the simple Virasoro \tsf{VOA} module with central charge $c$ and highest weight $h$ by $\mr{Vir}(c,h)$. Using the results of \cite{Goddard:1984vk} we find, for $k=1$: 
\begin{eqsp}\label{eq:su(2)-ising-hilbert-space}
    \mathscr{H}_{G/H}^{\mr{WZW}}&=\left(\mr{Vir}\left(\frac{1}{2},0\right)\otimes \wt{\mr{Vir}}\left(\frac{1}{2},0\right)\right)\otimes(\IC |0\rrangle\oplus  \IC|2\rrangle)
    \\
    &\oplus \left(\mr{Vir}\left(\frac{1}{2},\frac{1}{2}\right)\otimes \wt{\mr{Vir}}\left(\frac{1}{2},\frac{1}{2}\right)\right)\otimes (\IC|0\rrangle\oplus \IC|2\rrangle)
    \\
&\oplus \left(\mr{Vir}\left(\frac{1}{2},\frac{1}{16}\right)\otimes \wt{\mr{Vir}}\left(\frac{1}{2},\frac{1}{16}\right)\right)\otimes (\IC|1\rrangle\oplus \IC|1\rrangle)~,
\end{eqsp}
where $|i\rrangle$ denotes the Ishibashi state corresponding to the $\mr{SU(2)}_k$ representation labeled by $i$, and the third degeneracy factor $(\IC|1\rrangle\oplus \IC|1\rrangle)$ is not a typo. 
The topological degeneracy is 2 and we have 
\begin{eqsp}
    \mr{End}(B)\cong \IC[\IZ_2]~,
\end{eqsp}
where $\IZ_2$ is the common center of $G=\mr{SU}(2)\times \mr{SU}(2),H=\mr{SU(2)}$.

\paragraph{Example 2: $G=\mr{Spin}(4n)_1,H=\mr{Sp}(n-1)_1\times\mr{SU(2)}_1\times \mr{SU}(2)_n$}
In this example we will need to discuss the cases $n$ odd and $n$ even separately. Here we will 
consider $n$ odd while the case of $n$ even is discussed in section \ref{sec:min_mod_Spin_n_even} below. Both these cases will be discussed in more detail in \cite{MRS}. 

The \tsf{GKO CFT} for this coset is again the unitary minimal model with central charge $c_n=1-\frac{6}{(n+1)(n+2)}$, for both parities of $n$. The parity dichotomy on $n$ comes from the fact that only in the case that $n$ is odd can we consider  $H$ to be a subgroup of $G$. In the case of   $n$ even, there is a homomorphism of $H$ into $G$ but it has a kernel and only $H/\IZ_2$ embeds in $G$. Thus for $n$ even the subgroup of $G$ is $H/\IZ_2$ and since this group is not simply connected it goes beyond the $Z$-regular case. 

The branching functions have been worked out explicitly in \cite{Bardakci:1987ee,Altschuler:1987zb}, using which one can show that the topological degeneracy is given by the group algebra $\IC[\IZ_2\times\IZ_2]$ of the common center $\IZ_2\times \IZ_2$ for all $n$ odd. The $n=1$ case leads us to the Verlinde algebra $V_1(\mr{Spin}(4))$ since $\mr{Spin(4)\cong SU(2)\times SU(2)}$. So, we demonstrate the topological degeneracy for the case $n=3$.  We find that 
\begin{eqsp}\label{eq:spin-ising-Hilbert-space}
    \mathscr{H}_{G/H}^{\mr{WZW}}&\cong\mr{Vir}\!\left(\frac{7}{10},0\right)
 \otimes\widetilde{\mr{Vir}}\!\left(\frac{7}{10},0\right)
 \otimes
 \Big[
 \mathbb C|\Lambda_0,0,0\rrangle
 \oplus\mathbb C|\Lambda_2,1,3\rrangle
 \oplus\mathbb C|\Lambda_0,0,3\rrangle
 \oplus\mathbb C|\Lambda_2,1,0\rrangle
 \Big]
\\
&\oplus
\mr{Vir}\!\left(\frac{7}{10},\frac{3}{5}\right)
 \otimes\widetilde{\mr{Vir}}\!\left(\frac{7}{10},\frac{3}{5}\right)
 \otimes
 \Big[
 \mathbb C|\Lambda_0,0,2\rrangle
 \oplus\mathbb C|\Lambda_2,1,1\rrangle
 \oplus\mathbb C|\Lambda_0,0,1\rrangle
 \oplus\mathbb C|\Lambda_2,1,2\rrangle
 \Big]
\\
&\oplus
\mr{Vir}\!\left(\frac{7}{10},\frac{3}{2}\right)
 \otimes\widetilde{\mr{Vir}}\!\left(\frac{7}{10},\frac{3}{2}\right)
 \otimes
 \Big[
 \mathbb C|\Lambda_2,0,0\rrangle
 \oplus\mathbb C|\Lambda_0,1,3\rrangle
 \oplus\mathbb C|\Lambda_2,0,3\rrangle
 \oplus\mathbb C|\Lambda_0,1,0\rrangle
 \Big]
\\
&\oplus
\mr{Vir}\!\left(\frac{7}{10},\frac{1}{10}\right)
 \otimes\widetilde{\mr{Vir}}\!\left(\frac{7}{10},\frac{1}{10}\right)
 \otimes
 \Big[
 \mathbb C|\Lambda_2,0,2\rrangle
 \oplus\mathbb C|\Lambda_0,1,1\rrangle
 \oplus\mathbb C|\Lambda_2,0,1\rrangle
 \oplus\mathbb C|\Lambda_0,1,2\rrangle
 \Big]
\\
&\oplus
\mr{Vir}\!\left(\frac{7}{10},\frac{7}{16}\right)
 \otimes\widetilde{\mr{Vir}}\!\left(\frac{7}{10},\frac{7}{16}\right)
 \otimes
 \Big[
 \mathbb C|\Lambda_1,1,0\rrangle
 \oplus\mathbb C|\Lambda_1,0,3\rrangle
 \oplus\mathbb C|\Lambda_1,1,3\rrangle
 \oplus\mathbb C|\Lambda_1,0,0\rrangle
 \Big]
\\
&\oplus
\mr{Vir}\!\left(\frac{7}{10},\frac{3}{80}\right)
 \otimes\widetilde{\mr{Vir}}\!\left(\frac{7}{10},\frac{3}{80}\right)
 \otimes
 \Big[
 \mathbb C|\Lambda_1,1,2\rrangle
 \oplus\mathbb C|\Lambda_1,0,1\rrangle
 \oplus\mathbb C|\Lambda_1,1,1\rrangle
 \oplus\mathbb C|\Lambda_1,0,2\rrangle
 \Big]~,
\end{eqsp}
where $\Lambda_j,j=0,1,2$ are the affine fundamental weights of $\mr{Sp(2)_1}$ and $\IC|\Lambda_j,t,u\rrangle$ denotes the Ishibashi state corresponding to the 
integrable highest weight representation of 
$H=\mr{Sp}(2)_1\times\mr{SU(2)}_1\times \mr{SU}(2)_3$ where the highest weight vector has spin $j$ with 
$2j = t,u$ for the two $\mr{SU(2)}$ factors, respectively. Thus the  topological degeneracy \footnote{Note that there is a typo in \cite[Equation (5.20)]{Altschuler:1987zb}. The degeneracy there is $4l$ which should be 4. A detailed calculation will appear in \cite{MRS}.} is $4$,
\begin{eqsp}
    \mr{End}(B)\cong \IC[\IZ_2\times\IZ_2]~,
\end{eqsp}
and, again, $\IZ_2\times\IZ_2$ is the common center of $G=\mr{Spin}(12),H=\mr{Sp}(2)\times\mr{SU(2)}\times \mr{SU}(2)$.

\subsection{Topological Degeneracy And The  Equivariant $H/H$ Theory}\label{sec:equivariant_H/H}

As discussed in the introduction, the $H/H$-\tsf{WZW} model is known to be a 2d \tsf{TFT} 
\cite{SPIEGELGLAS199036,Spiegelglas:1991uc,Spiegelglas:1992jg,Witten:1991mm,Witten:1993xi}.
On the other hand, when $H=G$ the \tsf{GKO} theory is trivial, so then the mismatch between the 
theories is simply the topological $H/H$ model. 
%
%
One could guess that, more generally, the $H/H$ theory should be related to the  topological sector 
of the general $G/H$-\tsf{WZW} model. It turns out that such a relation exists, but it involves the equivariant $H/H$-\tsf{WZW} model as we explain below. We begin by briefly recalling the $H/H$-\tsf{WZW} model. 

\subsubsection{The $H/H$-\tsf{WZW} Model}

  The commutative Frobenius algebra corresponding to the $H/H$-\tsf{WZW} model is the \textit{Verlinde algebra} $V_{\wt\bk}(H)$ \cite{Witten:1993xi}. As a vector space
\begin{equation}
V_{\widetilde{\bk}}(H)=\mr{Span}_{\mathbb{C}}\left\{\CO_{\bmu}: \bmu \in P_{\widetilde{\bk}}^{+}(\fh)\right\}~.
\end{equation}
The multiplication on $V_{\widetilde{\bk}}(H)$ is given by
\begin{equation}\label{eq:prod_G/GTFT}
\CO_{\bmu_1} \cdot \CO_{\bmu_2}=\sum_{\bmu_3 \in P_{\wt{\bk}}^{+}(\fh)} \CN_{\bmu_1 \bmu_2}^{\fh, \bmu_3} \CO_{\bmu_3}~,
\end{equation}
where the $\CN_{\bmu_1 \bmu_2}^{\fh, \bmu_3} \in \mathds{Z}_{\geq 0}$ are the fusion rules. Note that the product is commutative and $\CO_0$ is the identity of the multiplication.  Then $V_{\wt{\bk}}(H)$ can be endowed with a Frobenius algebra structure by defining the Frobenius pairing to be
\begin{equation}\label{eq:frob_pair_H/H}
\left\langle\CO_{\bmu_1}, \CO_{\bmu_2}\right\rangle=\delta_{\bmu_2, \bmu_1^*}~,
\end{equation}
where $\bmu^*$ is the weight corresponding to the dual of the module $L_{\fh}(\wt{\bk},\hbmu)$.


\subsubsection{Gauged \tsf{WZW} Hilbert Space As Coupling Of \tsf{GKO} And Equivariant $H/H$}

We now observe that the   Hilbert space \eqref{eq:G/H_space_RepS} can also be interpreted as a coupling between $\mathscr{H}_{\fg/\fh}^{\mr{GKO}}$ and the $Z$-equivariant extension of the $H/H$-\tsf{WZW} model. Note that $Z\cong G_{\mr{id}}$   acts on   $P^+_{\widetilde \bk}(\fh)$ via $(J,J')\cdot \CO_{\bmu}=\CO_{J'\bmu}$.  The Hilbert space of the $Z$-equivariant $H/H$-\tsf{WZW} model can be worked out using the description of the \textit{algebra of little loops} of \cite{Moore:2006dw} 
as appropriate to a $Z$-equivariant 2d \tsf{TFT} based on  $P^+_{\widetilde \bk}(\fh)//Z$.
\footnote{The algebra of little loops is an example of a \emph{Turaev algebra} \cite{Moore:2006dw}, a concept derived from \cite{Turaev:1999yf}.}
 Explicitly, the Hilbert space of the $Z$-equivariant $H/H$ model on the circle  is: 
\begin{eqsp}\label{eq:Z-equi-HmodH}
    V_{\wt{\bk}}(H)_{Z}:=\bigoplus_{\bmu\in P^+_{\wt{\bk}}(\fh)}\C[\CS_{\bmu}]~,
\end{eqsp}
where $\CS_{\bmu}< G_{\mr{id}}$ is the stabilizer subgroup of $\bmu$. The summands in 
\eqref{eq:Z-equi-HmodH} are just the degeneracy spaces we observed in 
\eqref{eq:G/H_space_RepS}. In this sense we can say that  the Hilbert space \eqref{eq:G/H_space_RepS} is given by the \tsf{GKO} Hilbert space ``intertwined'' with the $Z$-equivariant $H/H$-\tsf{WZW} model. Our formula \eqref{eq:G/H_space_RepS} expresses this ``intertwining'' precisely, but a more conceptual understanding remains to be discovered.

\subsubsection{The Curious Special Case Of The $\mr{U(1)/U(1)}$ Theory }

As a parenthetical remark we comment on the  case $H=\mr{U(1)}$ (which is the Gaussian model with radius squared given by an integer). This is quite interesting because the conjugation action of $\mr{U(1)}$   on itself is trivial.  One might wonder how gauging a symmetry that acts trivially on the fields could possibly eliminate the physical degrees of freedom leaving ``only'' a finite dimensional space of topological modes. In general if a group $G$ acts trivially on the fields of a Lagrangian field theory $T$ we would expect that ``gauging the $G$-symmetry of $T$'' would simply produce the direct product of $G$-Yang-Mills and $T$. That is not the case here because - in general, and in this particular example - the gauged \tsf{WZ}-term cannot be deduced from gauging the \tsf{WZ}-term in the absence of gauge fields. Whether one follows the approach using equivariant cohomology as in \cite{Witten:1991mm,Figueroa-OFarrill:1994uwr,Figueroa-OFarrill:1994vwl}
  or the more general approach using differential cohomology one finds that even though the \tsf{WZ}-term is zero in the absence of gauge fields the gauged \tsf{WZ}-term is nontrivial. In the case of $H=\mr{U(1)}$ there is a simple and explicit expression for the gauged \tsf{WZ}-term in differential 
  cohomology:   The \tsf{WZW} field determines an element of $\check H^1(\Sigma)$, whereas 
the $\mr{U(1)}$ gauge field determines an element of $\check H^2(\Sigma)$. The gauged \tsf{WZ}-term is 
simply given by the canonical pairing of multiplying and integrating in differential cohomology: 
\be\label{eq:DiffCohoPairing}
\check H^1(\Sigma) \times \check H^2(\Sigma) \to \mr{U(1)}~. 
\ee
When the winding numbers of the \tsf{WZW} field $g(z,\bar z)$ are zero, so that we may write 
$g = \exp[ \im\, \phi(z,\bar z)]$ for a well-defined scalar field $\phi$ on $\Sigma$, we can write
the pairing in \eqref{eq:DiffCohoPairing}  as 
$\exp[ \frac{\im}{2\pi}  \int_{\Sigma}  \phi F ]$. Then   the sum over gauge bundles kills the momentum modes. When the 
gauge field is topologically trivial so we can write $F=\rd A$ for a globally well-defined 1-form $A$ on $\Sigma$, we can write
equation \eqref{eq:DiffCohoPairing}  as 
 $\exp[ \frac{1}{2\pi} \int g^{-1} \rd g \wedge A] $. In this case the integral over the flat gauge fields kills the winding modes. Then the $\mr{U(1)}$ ghosts kill the oscillator degrees of freedom in $\phi$.
 A detailed study using the Hamiltonian approach makes clear that the level $k$ $\mr{U(1)/U(1)}$ model is indeed the Verlinde algebra. A detailed account will appear in \cite{MRS}.  

The above result fits together nicely with $T$-duality (see \cite{Giveon:1994fu} for a review). On-shell we can write the \tsf{WZW} field 
as $g(z,\bar z) = \exp[ \im (\phi_L(z) + \phi_R(\bar z)]$. The conjugation action by $e^{\im \alpha} \in \mr{U(1)}$ can be viewed as 
\be 
\phi_L(z) \rightarrow  \phi_L(z) + \alpha~, \qquad \qquad \phi_R(\bar z) \rightarrow \phi_R(\bar z) - \alpha~ 
\ee
and indeed acts trivially on $g(z,\bar z)$. However,   the $T$-dual field $\tilde \phi := \phi_L(z) - \phi_R(\bar z)$ shifts by $2\alpha$ under the $\mr{U(1)}$ action so that it is no surprise that gauging this action leads to a 2d \tsf{TFT}. Thus our discussion is also related to the path integral derivation of $T$-duality discussed in \cite{Rocek:1991ps}.

\section{Conjectured Extension To All Pairs Of Compact Lie Groups $H< G$}\label{sec:other_examples}

We have phrased Theorem   \ref{thm:main_thm} in such a way that the statement makes sense
with much weaker hypotheses. Indeed, it is phrased so that it makes sense for any 
pair of compact Lie groups $H< G$. This allows us to state our: 

\begin{conj}
Let $G,H$ be any pair of compact  Lie groups with  $H< G$. Then we have 
\begin{eqsp}
    \mr{WZW}(G,H;k)\cong \mr{GKO}(G,H;k)\,\wt{\otimes}\,\mr{End}(B)~,  
\end{eqsp}
where $B$ is an algebra object in the category $\CC(G;k)\boxtimes_{\mr{D}}\CC^{\mr{opp}}(H;\tilde k)$ generalizing \eqref{eq:B_def_affine} and \eqref{eq:comm_cat_affine}. Here  $\CC(G;k)$ is the modular tensor category of $V(G;k)$-modules and $\CC(H;\tilde k)$ is defined similarly.
\end{conj}

We will now give several examples that illustrate this conjecture. The main thrust of our examples is to show that the result extends beyond the $Z$-regular case with $H<G$ connected, simply connected, and semisimple. 

\subsection{Parafermions: $G=\mr{SU}(2)_k,H=\mr{U}(1)_k$.}\label{sec:parafermions}

This is an interesting case because while the coset is $Z$-regular, $H$ is not simply connected. 
Let us label the weights of $\wh{\fu}(1)_k$ by $n=-k+1,\dots,0,\dots,k$. We have the decomposition \cite{DiFrancesco:1997nk}:
\begin{eqsp}\label{eq:branch_SU(2)_U(1)}
    L_{\fs\fu(2)}(k,i)\cong\bigoplus_{\substack{n=-k+1\\i+n\equiv 0\bmod 2}}^{k}L_{\fs\fu(2)/\fu(1)}(i,n)\otimes L_{\fu(1)}(k,n)~. 
\end{eqsp}
The \tsf{GKO VOA} $L_{\fs\fu(2)/\fu(1)}(0,0)$ is a $c_k=\frac{2(k-1)}{k+2}$ \tsf{VOA}. Using the decomposition \eqref{eq:branch_SU(2)_U(1)}, one can compute $\mathscr{H}_{\mr{SU(2)/U(1)}}^{\mr{WZW}}$ and we find that the topological degeneracy 
\footnote{Since $\pi_0(L\mr{U(1)})\cong\pi_1(\mr{U(1)})\cong \IZ$, there are more constraints from Gauss' law. As we show in \cite{MRS}, the result is unchanged and the topological degeneracy is still 2.}
is $2=|\IZ_2|$ where $\IZ_2$ is the common center of $\mr{SU(2)}$ and $\mr{U(1)}$. The algebra object $B$ is given by 
\begin{eqsp}
    B=L_{\fs\fu(2)}(k,0)\boxtimes_{\mr{D}}L^{\mr{opp}}_{\fu(1)}(k,0)\oplus  L_{\fs\fu(2)}(k,k)\boxtimes_{\mr{D}}L^{\mr{opp}}_{\fu(1)}(k,k)~,
\end{eqsp}
and 
\begin{eqsp}
    \mr{End}(B)\cong \IC[\IZ_2]~.
\end{eqsp}
%
%

\subsection{Conformal Embeddings}

One  prominent class of examples where the usual selection rules and field identifications do not work are conformal embeddings.  
The isomorphism \eqref{eq:Main-EndB-statement} continues to hold for conformal embeddings. Indeed, for conformal embeddings with $G,H$ simply connected, we have 
\begin{eqsp}
\mathscr{H}^{\mr{WZW}}_{G/H}:=\bigoplus_{(\blm,\bmu)\in\CE} \left(\C^{n_{(\hblm,\hbmu)}} \otimes \wt{\C}^{n_{(\hblm,\hbmu)}}\right)\otimes\IC|\bmu\rrangle~,    
\end{eqsp}
where the integers $n_{(\hblm, \hbmu)}$ is defined in \eqref{eq:nlm_def}. We thus see that 
\begin{eqsp}
\mathscr{H}^{\mr{WZW}}_{G/H}\cong \mr{End}(B)~. 
\end{eqsp}

\subsubsection{$G/G$-\tsf{WZW} Model}
The case $H=G$ is a special case of a conformal embedding. In the $G/G$ case with $G$ simply connected, we have 
\begin{eqsp}
    B=\bigoplus_{\blm\in P^+_{\bk}(\fg)}L_{\fg}(\bk,\hblm)\boxtimes_{\mr{D}}L^{\mr{opp}}_{\fg}(\bk,\hblm)~.
\end{eqsp}
Working through the construction of the multiplication and comultiplication map on $B$ described in \cite{Frohlich:2003hm} and using \eqref{eq:mult_EndB}, we find that 
\begin{eqsp}
    f_{(\hblm,\hblm)}\cdot f_{(\hblm',\hblm')}=\sum_{\blm''\in P^+_{\bk}(\fg)}\CN_{\hblm\hblm'}^{\fg,\hblm''}f_{(\hblm'',\hblm'')}~,
\end{eqsp}
where $f_{(\hblm,\hblm)}$ is defined in analogy to \eqref{eq:fJJ'_def} and \eqref{eq:iJJ'rJJ'_def}. 
This agrees with the multiplication on $V_{\bk}(G)$. The Frobenius trace on $\mr{End}(B)$ can also be calculated to be
\begin{eqsp}
    \mr{Tr}_B(f_{\hblm,\hblm})=\delta_{\hblm,0}\varepsilon_B\circ\eta_B~.
\end{eqsp}
One can rescale the counit $\eta_B$ such that 
\begin{eqsp}
\mr{Tr}_B(f_{\hblm,\hblm})=\delta_{\hblm,0}~.    
\end{eqsp}
Thus we have the isomorphism 
\begin{eqsp}
V_{\bk}(G)\cong \mr{End}(B)~,    
\end{eqsp}
of the Verlinde algebra describing the \tsf{TFT} structure of the $G/G$-\tsf{WZW} model with $\mr{End}(B)$ 
as expected.

\subsubsection{Example: $G=\mr{Spin}(N^2 -1)_1$, $H=(\mr{SU}(N)/\Gamma)_N$}

This conformal embedding has been studied in  \cite{Komargodski:2020mxz} where, among many other things, it was noted that for $H$ to be a subgroup we must quotient 
$\mr{SU}(N)$ by the finite group  
%
%
\begin{eqsp}
    \Gamma=\begin{cases}
        \IZ_N,&N~\text{odd},
        \\
        \IZ_{N/2},&N~\text{even}.
    \end{cases}
\end{eqsp}
Since $H$ is not simply connected from $N>2$, we need the generalization of \eqref{eq:GmodH-WZW-space}. Working out the algebra object $B$ in this case, one can show that for $N>2$ \cite{MRS}
\footnote{For $N=2$ our result agrees with \cite{Komargodski:2020mxz} but for $N>2$ our result
differs from \cite{Komargodski:2020mxz}. The discrepancy arises because $H$ is not simply connected. If one ignores subtleties arising from the fact that $H$ is not simply connected 
the inequality \eqref{eq:EndB-inequality-KORS} is replaced an equality, as originally stated in 
$v1$ of this paper. }
%
%
\begin{eqsp}\label{eq:EndB-inequality-KORS}
\mr{dim}\,\mathscr{H}_{G/H}^{\mr{WZW}}=\mr{dim}\,\mr{End}(B)<3\times 2^{N-2}~.    
\end{eqsp}
%
%
Our results provide a commutative Frobenius algebra structure on $\mr{End}(B)$, thus identifying it with a 2d \tsf{TFT}.  For $N=3$, one can compute $\mr{dim}\,\mr{End}(B)=4$ and also explicitly describe the Frobenius algebra structure on $\mr{End}(B)$. The details will appear in \cite{MRS}.
It would be very interesting to understand the 
\underline{commutative} Frobenius algebra structure on $\mr{End}(B)$ for $N>3$. 

\subsubsection{Example: $G=(\mr{G}_2)_1,H=\mr{SU(2)_1\times\mr{SU(2)_3}}$.}
Using the decomposition of $(\wh{\fg}_2)_1$ representations into $\wh{\fs\fu}(2)_1\oplus\wh{\mathfrak{su}}(2)_3$ representations, we find
\begin{eqsp}
    \mathscr{H}^{\mr{WZW}}_{(\mr{G}_2)_1/\mr{SU(2)_1\times\mr{SU(2)_3}}}&\cong  \IC|0,0\rrangle\oplus \IC|1,3\rrangle\oplus \IC|0,2\rrangle\oplus \IC|1,1\rrangle~.
\end{eqsp}
Thus, the topological degeneracy is 4. The algebra object $B$ is given by 
\begin{eqsp}
    B&=L_{\fg_2}(1,\textbf{1}) \boxtimes_{\mr{D}} L^{\mr{opp}}_{\mf{su}(2)\oplus \mf{su}(2)}((1,3),(0,0)) \oplus L_{\fg_2}(1,\textbf{1}) \boxtimes_{\mr{D}}L^{\mr{opp}}_{\mf{su}(2)\oplus \mf{su}(2)}((1,3),(1,3))\\&\oplus L_{\fg_2}(1,\textbf{7}) \boxtimes_{\mr{D}} L^{\mr{opp}}_{\mf{su}(2)\oplus \mf{su}(2)}((1,3),(0,2)) L_{\fg_2}(1,\textbf{7}) \boxtimes_{\mr{D}}L^{\mr{opp}}_{\mf{su}(2)\oplus \mf{su}(2)}((1,3),(1,1))~ ,
\end{eqsp}
where we are denoting the dominant integral weights of $(\wh{\fg_2})_1$ by the dimension of the corresponding finite-dimensional representation in bold. 
Thus, $\mr{End}(B)$ is also 4-dimensional. Checking the Frobenius algebra structure on $\mr{End}(B)$, we find the following isomorphism
\begin{eqsp}
    \mathscr{H}^{\mr{WZW}}_{(\mr{G}_2)_1/\mr{SU(2)_1\times \mr{SU(2)_3}}}\cong V_{1}(\mr{SU(2)})\otimes V_{1}(\mr{G_2})~.
\end{eqsp}
This isomorphism was conjectured in \cite{Cordova:2023jip}.

\subsection{$G=\mr{Spin}(4n)_1,H=(\mr{Sp}(n-1)_1\times\mr{SU(2)}_1\times \mr{SU}(2)_n)/\IZ_2$ }\label{sec:min_mod_Spin_n_even}
Now we turn to the case where $n$ is even. In this case there is no injective 
homomorphism of $\mr{Sp}(n-1)_1\times\mr{SU(2)}_1\times \mr{SU}(2)_n$ into $G$. Since we always assume $H$ is a subgroup of $G$ we must take 
$H=(\mr{Sp}(n-1)_1\times\mr{SU(2)}_1\times \mr{SU}(2)_n)/\IZ_2$. 
In this case, $H$ is not simply connected. Disconnected components of $LH$ lead to further identifications among Ishibashi state. The resulting topological degeneracy is 2 \cite{MRS}. One can also show that $\mr{End}(B)$ is 2-dimensional \cite{MRS}. We explicitly write the answer for $n=2$:
\begin{eqsp}\label{eq:spin-ising-Hilbert-space}
    \mathscr{H}_{G/H}^{\mr{WZW}}&=\left(\mr{Vir}\left(\frac{1}{2},0\right)\otimes \wt{\mr{Vir}}\left(\frac{1}{2},0\right)\right)\otimes \Big[\IC(|0,0,0\rrangle+|1,1,2\rrangle)\oplus\IC(|0,0,2\rrangle+|1,1,0\rrangle)\Big]
    \\
    &\oplus \left(\mr{Vir}\left(\frac{1}{2},\frac{1}{2}\right)\otimes \wt{\mr{Vir}}\left(\frac{1}{2},\frac{1}{2}\right)\right)\otimes \Big[\IC(|1,1,0\rrangle+|0,0,2\rrangle)\oplus\IC(|0,0,0\rrangle+|1,1,2\rrangle)\Big]
    \\
&\oplus \left(\mr{Vir}\left(\frac{1}{2},\frac{1}{16}\right)\otimes \wt{\mr{Vir}}\left(\frac{1}{2},\frac{1}{16}\right)\right)\otimes \Big[\IC(|0,1,1\rrangle+|1,0,1\rrangle)\oplus\IC(|1,0,1\rrangle+|0,1,1\rrangle)\Big]~,
\end{eqsp}
where $\C|s, t, u\rrangle$ denotes the Ishibashi state corresponding to the integrable highest weight
representation of $H = \mr{Sp}(1)_1 \times \mr{SU(2)_1 \times SU(2)_2}$ where the highest weight vector has spin
$j$ with $2j = s, t, u,$ respectively

\subsection{A Maverick Example: $G=\mr{SU(3)_2},H=\mr{SO(3)_8}$}\label{sec:maverick}
The \tsf{GKO} \tsf{VOA} in this case is the $\mr{W}_3$ algebra with central charge $c=\frac{4}{5}$. The \tsf{GKO} theory is thus the three-state-Potts-model \cite{dunbar1993characters,Dunbar:1993hr}. Thus, the \tsf{GKO VOA} has six irreducible modules, which we denote by $V_{\mathbb{I}},V_{\epsilon},V_{\psi_1},V_{\psi_2},V_{\sigma_1},V_{\sigma_2}$.Since $H$ is not simply connected, we need to use the generalization of \eqref{eq:GmodH-WZW-space}. The gauged \tsf{WZW} Hilbert space is given by \cite{MRS}
\begin{eqsp}\label{eq:su(3)/su(2)WZWHil}
    \mathscr{H}^{\mr{WZW}}_{\mr{SU(3)_2/SO(3)_8}}\cong (V_\mathbb{I}\otimes \wt{V}_\mathbb{I})&\otimes(\IC(|0\rrangle+|8\rrangle)\oplus\IC|4\rrangle)
    \\\oplus (V_\epsilon\otimes \wt{V}_\epsilon)&\otimes(\IC|4\rrangle\oplus\IC(|6\rrangle+|2\rrangle))  
    \\\oplus (V_{\psi_1}\otimes \wt{V}_{\psi_1})&\otimes(\IC(|0\rrangle+|8\rrangle)\oplus\IC|4\rrangle)\\\oplus (V_{\psi_2}\otimes \wt{V}_{\psi_2})&\otimes(\IC(|0\rrangle+|8\rrangle)\oplus\IC|4\rrangle)\\\oplus (V_{\sigma_1}\otimes \wt{V}_{\sigma_1})&\otimes(\IC|4\rrangle\oplus\IC(|6\rrangle+|2\rrangle))\\\oplus (V_{\sigma_2}\otimes \wt{V}_{\sigma_2})&\otimes(\IC|4\rrangle\oplus\IC(|6\rrangle+|2\rrangle))~.    
\end{eqsp}
We see that the topological degeneracy is 2. The algebra object $B$ is given by 
\begin{eqsp}
    B=\left(L_{\fs\fu(3)}(2,\textbf{1})\boxtimes_{\mr{D}}\left(L_{\fs\fu(2)}(8,0)\oplus L_{\fs\fu(2)}(8,8)\right)^{\mr{opp}}\right) \oplus \left(L_{\fs\fu(3)}(2,\textbf{8})\boxtimes_{\mr{D}}L^{\mr{opp}}_{\fs\fu(2)}(8,4)\right)~, 
\end{eqsp}
where we are denoting the dominant integral weights of $\wh{\fs\fu}(3)_2$ by the dimension of the corresponding finite-dimensional representation in bold. We see that $\mr{End}(B)$ is also two dimensional. Again, the topological sector tensored with the vacuum module $V_{\mathbb{I}}\otimes\wt{V}_{\mathbb{I}}$ is isomorphic to $\mr{End}(B)$ while the topological sector of other summands in the decomposition \eqref{eq:su(3)/su(2)WZWHil} involve 
rank 1 free modules for $\mr{End}(B)$ \cite{MRS}. 
%

%
%

We now turn to the implications of our results.

\section{Applications}
\subsection{String Theory Implications}\label{sec:implications}

Our results might have some applications to string theory. Coset models in string theory can arise in two ways. First, coset models with compact groups can be used as building blocks for string compactifications. Second, coset models involving non-compact groups, in particular $\mr{SL}(2,\IR)$, have been employed for discussing various types of black holes as well as cosmologies \cite{Elitzur:1990ubs,Mandal:1991tz,Witten:1991yr,Elitzur:2002rt,Nappi:1992kv,Kounnas:1992wc,Giveon:2003gb} within the context of string theory. 

In the first case our results indicate that if one uses the gauged \tsf{WZW} model rather than the \tsf{GKO} model to construct the internal dimensions then the string theory will have interesting topological symmetry. Equation \eqref{eq:RescaleTorus} indicates that there will be a finite renormalization of the squared string coupling by $\mr{dim}\, \mr{End}(B)$. Further work is needed to confirm this.

The second case involves \tsf{WZW} models for noncompact groups. These are not \tsf{RCFT}s and many new issues arise. It should be very interesting to see if and how the phenomenon we have discussed here extends to such models. It might have interesting implications for the study of the entropy of models of black holes in string theory. Again, there are potentially important renormalizations of the string coupling constant. Moreover, topological degrees of freedom in a black hole might signal the existence of remnants. 

As a final comment we would note that if  we simply assume that our results apply to the $\mr{SL}(2,\IR)_k/\mr{U(1)}_{4k}$ models then there is an interesting contrast between our results and the naive expectation that the topological degrees of freedom are an overall product with the $H/H$ model. In the latter case the topological degeneracy leads to a renormalization of string amplitudes at genus $g$ by a factor $(8k)^{g}$ and therefore the string coupling would have been renormalized by $g_{str} \to g_{str} \sqrt{8k}$. Thus, the semiclassical limit would have involved a subtle order of limits issue in string perturbation theory. Our results (if applicable to the $\mr{SL}(2,\IR)_k/\mr{U(1)}_{4k}$ model) indicate that  in fact the string coupling only gets a finite renormalization, independent of $k$. 

\subsection{Infrared Phases Of 2d Yang-Mills (\tsf{YM}) With Matter}\label{sec:2dQCD}

For a compact Lie group $G$, we can consider the 2d Yang-Mills of left and right chiral quarks in the representation $\mf{R}$ of $G$.
\footnote{In general, we can consider the left and right chiral quarks in different representations $\mf{R}_{\ell}$ and $\mf{R}_r$ respectively chosen appropriately such that the theory is anomaly-free. We will restrict to the case when the left and right chiral quarks are in the same representation.}
Nonabelian bosonization \cite{Witten:1983ar} then states that this 2d \tsf{YM} with matter is equivalent to the gauged \tsf{WZW} model $\mr{WZW}(\mr{Spin}(\mr{dim}\,\mf{R}),G;1)$ \textit{with} the Maxwell term. In making this correspondence 
it is necessary to gauge  $(-1)^{\mr{F}}$,  as explained in papers going back to 
\cite{Alvarez-Gaume:1986nqf,Alvarez-Gaume:1987wwg,Elitzur:1986ye}. 
As we have seen in section \ref{sec:IR_G/HWZW}, taking the $e^2\to\infty$ limit results in the gauged \tsf{WZW} model. Hence the gauged \tsf{WZW} model describes the \tsf{IR} limit of this (bosonized) 2d \tsf{YM} with matter \cite{Delmastro:2021otj,Komargodski:2020mxz}. Moreover, 
\cite{Komargodski:2020mxz} studied the symmetries of 2d Yang-Mills with adjoint matter. See also \cite{Nguyen:2021naa,Cherman:2019hbq,Huang:2021zvu}. 
These symmetries are expected to be \tsf{RG} invariants.
This motivates us to study the ``zero-form'' and ``one-form'' symmetries of the gauged \tsf{WZW} model. The results of this paper shed light on those symmetries. \par

%
%
%
%
%

We identify the ``zero-form symmetries'' with the fusion category of topological defect lines (\tsf{TDL}s) \cite{Chang:2018iay}. In general the category of \tsf{TDL}s has infinitely many simple objects. For \tsf{RCFT}s, there is a subset of \tsf{TDL}s, called \textit{Verlinde lines}, which commute with all the generators of the chiral algebra. For the gauged \tsf{WZW} model, we have the usual Verlinde lines which act on $\mathscr{H}_{G/H}^{\mr{GKO}}$. The category of these Verlinde lines is the \tsf{MTC} $\CC(\fg,\fh;\bk)$. There are other Verlinde lines which act on the topological sector $\mr{End}(B)$ of the gauged \tsf{WZW} model. We leave a detailed investigation of the precise structure of the Verlinde lines in gauged \tsf{WZW} model for a different occasion.    

%
%
%
%
%
%
%
%

We now turn to the ``one-form symmetries.'' Since we are discussing an $H$-gauge theory we can automatically consider the symmetries associated with the center of $H$. Turning on a ``background one-form symmetry gauge field'' simply means that we consider the theory formulated with an 
$\tilde H$ gauge bundle $\tilde P$ where $\tilde H = H/\Gamma$ and $\Gamma$ is a subgroup of $Z(H)$. 
The ``background one-form symmetry gauge field'' is nontrivial if the $\tilde H$-bundle $\tilde P$ does not admit a reduction of structure group to $H$. The obstruction to doing so is measured by $w_2(\tilde P) \in H^2(\Sigma; Z(H))$. The \tsf{WZW} field must be a section of the associated bundle with fiber $G$ and adjoint action of $\tilde H$. The adjoint action of $\tilde H$ on $G$ will only be well-defined if $\Gamma \subset Z(G)$. The maximal choice for $\Gamma$ is the common center $Z$. Thus, we recover the expected result that the group $Z$ is isomorphic to a  group of one-form symmetries. 

It is worth noting that one-form symmetries are often identified with invertible topological operators. 
\footnote{In general one views one-form symmetries as acting on 1d defects via the linking sphere. In 2d the linking sphere is the zero-dimensional sphere and requires a \underline{pair} of topological operators.}
We will adopt this viewpoint. Every element of $\mr{End}(B)$ defines a topological point operator which acts on $\mathscr{H}_{G/H}^{\mr{WZW}}$ as
\begin{eqsp}
    f\cdot (\mr{v}\otimes g)=\mr{v}\otimes(f\cdot g)~,\quad \mr{v}\in \mathscr{H}_{G/H}^{\mr{GKO}},f,g\in\mr{End}(B)~, 
\end{eqsp}
where $f\cdot g$ makes use of the multiplication defined in  \eqref{eq:mult_EndB}.
Let us denote the group of invertible topological point operators by $\Gamma_B$. 
In the case where $\tsf{End}(B) = \IC[Z]$ we can be fairly explicit about the group of invertible topological point operators: By the Peter-Weyl theorem 
\begin{eqsp}
\Gamma_B:=\IC[Z]^* \cong \prod_{\mr{Irrep}(Z)} \IC^*~.    
\end{eqsp}
The explicit isomorphism can be written as follows. Let $\mu$ run over the irreps of $Z$ and let $\chi^\mu$ be the characters in these (one-dimensional) irreps. The orthogonality relations for characters imply that 
\begin{eqsp}
e_\mu = \frac{1}{\vert Z\vert} \sum_{z\in Z} \chi^\mu(z^{-1}) z ~,    
\end{eqsp}
satisfy 
\begin{eqsp}
e_\mu e_\nu = \delta_{\mu,\nu} e_\nu~.    
\end{eqsp}
Note that   $Z$ only forms a proper subgroup of the group of invertible topological point operators.
\par   The space of vacua $\mr{End}(B)$ is decomposed into
``universes'' (in the terminology of \cite{Komargodski:2020mxz}) which are the superselection sectors determined by diagonalization of the action of $\Gamma_B$. Further comparison to the results of  \cite{Komargodski:2020mxz} will be discussed in \cite{MRS}.

\cleardoublepage
\phantomsection
\addcontentsline{toc}{section}{References}
\bibliographystyle{ytamsalpha} 
\bibliography{wzw-tft}

@book{FLM1988,
  title={Vertex Operator Algebras and the Monster},
  author={Frenkel, I. and Lepowsky, J. and Meurman, A.},
  isbn={9780080874548},
  series={Pure and Applied Mathematics},
  url={https://books.google.com/books?id=yOC9OFC1YSUC},
  year={1989},
  publisher={Academic Press}
}

@article{Chang:2018iay,
    author = "Chang, Chi-Ming and Lin, Ying-Hsuan and Shao, Shu-Heng and Wang, Yifan and Yin, Xi",
    title = "{Topological Defect Lines and Renormalization Group Flows in Two Dimensions}",
    eprint = "1802.04445",
    archivePrefix = "arXiv",
    primaryClass = "hep-th",
    reportNumber = "CALT-TH-2017-067, CALT-TH 2017-067, PUPT-2546",
    doi = "10.1007/JHEP01(2019)026",
    journal = "JHEP",
    volume = "01",
    pages = "026",
    year = "2019"
}

@book{PressleySegal1986,
  author    = {Pressley, Andrew and Segal, Graeme},
  title     = {Loop Groups},
  series    = {Oxford Mathematical Monographs},
  publisher = {Oxford University Press},
  address   = {Oxford},
  year      = {1986},
  isbn      = {9780198535356}
}

@article{Fuchs:1990wb,
    author = "Fuchs, Jurgen",
    title = "{Simple WZW currents}",
    reportNumber = "PRINT-90-0521 (NIKHEF), NIKHEF-H-90-14",
    doi = "10.1007/BF02100029",
    journal = "Commun. Math. Phys.",
    volume = "136",
    pages = "345--356",
    year = "1991"
}

@article{Witten:1983ar,
    author = "Witten, Edward",
    editor = "Stone, M.",
    title = "{Nonabelian Bosonization in Two-Dimensions}",
    reportNumber = "PRINT-83-0934 (PRINCETON)",
    doi = "10.1007/BF01215276",
    journal = "Commun. Math. Phys.",
    volume = "92",
    pages = "455--472",
    year = "1984"
}

@article{Witten:1993xi,
    author = "Witten, Edward",
    title = "{The Verlinde algebra and the cohomology of the Grassmannian}",
    eprint = "hep-th/9312104",
    archivePrefix = "arXiv",
    reportNumber = "IASSNS-HEP-93-41",
    month = "12",
    year = "1993"
}

@article{Witten:1991mm,
    author = "Witten, Edward",
    title = "{On Holomorphic factorization of WZW and coset models}",
    reportNumber = "IASSNS-HEP-91-25",
    doi = "10.1007/BF02099196",
    journal = "Commun. Math. Phys.",
    volume = "144",
    pages = "189--212",
    year = "1992"
}

@article{Elitzur:1990ubs,
    author = "Elitzur, S. and Forge, A. and Rabinovici, E.",
    title = "{Some global aspects of string compactifications}",
    reportNumber = "RI-143-90, IMPERIAL-TP-90-91-06",
    doi = "10.1016/0550-3213(91)90073-7",
    journal = "Nucl. Phys. B",
    volume = "359",
    pages = "581--610",
    year = "1991"
}

@article{witten1989quantum,
  title={Quantum field theory and the {Jones} polynomial},
  author={Witten, Edward},
  journal={Communications in Mathematical Physics},
  volume={121},
  number={3},
  pages={351--399},
  year={1989},
  publisher={Springer}
}

@article{JL1,
  title={The commutant of $\mathrm{L}_{\widehat{\mathfrak{sl}}_2}(n, 0)$ in the vertex operator algebra $\mathrm{L}_{\widehat{\mathfrak{sl}}_2}(1,0)^{\otimes n}$},
  author={Jiang, Cuipo and Lin, Zongzhu},
eprint = "1311.0608",
    archivePrefix = "arXiv",
  year={2013}
}

@article{Moore:1988ss,
    author = "Moore, Gregory W. and Seiberg, Nathan",
    title = "{Naturality in Conformal Field Theory}",
    reportNumber = "IASSNS-HEP/88/31",
    doi = "10.1016/0550-3213(89)90511-7",
    journal = "Nucl. Phys. B",
    volume = "313",
    pages = "16--40",
    year = "1989"
}

@article{Dunbar:1993hr,
    author = "Dunbar, David C. and Joshi, Keith G.",
    title = "{Maverick examples of coset conformal field theories}",
    eprint = "hep-th/9309093",
    archivePrefix = "arXiv",
    reportNumber = "LTH-93-310, SWAT-93-07",
    doi = "10.1142/S0217732393003196",
    journal = "Mod. Phys. Lett. A",
    volume = "8",
    pages = "2803--2814",
    year = "1993"
}

@article{dunbar1993characters,
  title={Characters for coset conformal field theories and maverick examples},
  author={Dunbar, David C and Joshi, Keith G},
  journal={International Journal of Modern Physics A},
  volume={8},
  number={23},
  pages={4103--4121},
  year={1993},
  publisher={World Scientific}
}

@article{Fuchs:1995tq,
    author = "Fuchs, Jurgen and Schellekens, Bert and Schweigert, Christoph",
    title = "{The resolution of field identification fixed points in diagonal coset theories}",
    eprint = "hep-th/9509105",
    archivePrefix = "arXiv",
    reportNumber = "NIKHEF-95-052, DESY-95-173",
    doi = "10.1016/0550-3213(95)00623-0",
    journal = "Nucl. Phys. B",
    volume = "461",
    pages = "371--406",
    year = "1996"
}

@inproceedings{Fuchs:1996rq,
    author = "Fuchs, Jurgen and Schellekens, Bert and Schweigert, Christoph",
    title = "{Fixed point resolution in conformal field theory}",
    booktitle = "{21st International Colloquium on Group Theoretical Methods in Physics}",
    eprint = "hep-th/9612093",
    archivePrefix = "arXiv",
    reportNumber = "DESY-96-251, NIKHEF-96-032, UCB-PTH-96-57, LBL-39689, LBNL-39689",
    month = "12",
    year = "1996"
}

@inproceedings{Moore:2025tmt,
    author = "Moore, Gregory W. and Saxena, Vivek and Freed, with an appendix by Daniel S.",
    title = "{TASI Lectures On Topological Field Theories And Differential Cohomology}",
    booktitle = "{Theoretical Advanced Study Institute in Elementary Particle Physics 2023}: {Aspects of Symmetry}",
    eprint = "2510.07408",
    archivePrefix = "arXiv",
    primaryClass = "hep-th",
    reportNumber = "YITP-SB-2025-04",
    month = "10",
    year = "2025"
}

@article{Moore:1989yh,
    author = "Moore, Gregory W. and Seiberg, Nathan",
    title = "{Taming the Conformal Zoo}",
    reportNumber = "IASSNS-HEP-89/6",
    doi = "10.1016/0370-2693(89)90897-6",
    journal = "Phys. Lett. B",
    volume = "220",
    pages = "422--430",
    year = "1989"
}

@article{gepner1989field,
  title={Field identification in coset conformal field theories},
  author={Gepner, Doron},
  journal={Physics Letters B},
  volume={222},
  number={2},
  pages={207--212},
  year={1989},
  publisher={Elsevier}
}

@article{Goddard:1984vk,
    author = "Goddard, P. and Kent, A. and Olive, David I.",
    title = "{Virasoro Algebras and Coset Space Models}",
    reportNumber = "DAMTP-84-22",
    doi = "10.1016/0370-2693(85)91145-1",
    journal = "Phys. Lett. B",
    volume = "152",
    pages = "88--92",
    year = "1985"
}

@article{Komargodski:2020mxz,
    author = "Komargodski, Zohar and Ohmori, Kantaro and Roumpedakis, Konstantinos and Seifnashri, Sahand",
    title = "{Symmetries and strings of adjoint QCD$_{2}$}",
    eprint = "2008.07567",
    archivePrefix = "arXiv",
    primaryClass = "hep-th",
    reportNumber = "YITP-SB-20-28",
    doi = "10.1007/JHEP03(2021)103",
    journal = "JHEP",
    volume = "03",
    pages = "103",
    year = "2021"
}

@article{JL2,
  title={Tensor decomposition, parafermions, level-rank duality, and reciprocity law for vertex operator algebras},
  author={Jiang, Cuipo and Lin, Zongzhu},
eprint = "1406.4191",
    archivePrefix = "arXiv",
  year={2014}
}

@article{ALY1,
  author = {Arakawa, Tomoyuki and Lam, Ching Hung and Yamada, Hiromichi},
  title = {Zhu’s Algebra, {$C_2$}-Algebra and {$C_2$}-Cofiniteness of Parafermion Vertex Operator Algebras},
  journal = {Advances in Mathematics},
  volume = {264},
  pages = {261–295},
  year = {2014}
}

@article{ALY2,
  author = {Arakawa, Tomoyuki and Lam, Ching Hung and Yamada, Hiromichi},
  title = {Parafermion Vertex Operator Algebras and {W}-Algebras},
  journal = {Transactions of the American Mathematical Society},
  volume = {371},
  number = {7},
  pages = {4277–4301},
  year = {2019}
}

@article{DW2,
  author = {Dong, Chongying and Wang, Qing},
  title = {On {$C_2$}-cofiniteness of parafermion vertex operator algebras},
  journal = {Journal of Algebra},
  volume = {328},
  pages = {420–431},
  year = {2011}
}

@article{Gawedzki:1988nj,
    author = "Gawedzki, K. and Kupiainen, A.",
    title = "{Coset Construction from Functional Integrals}",
    reportNumber = "HU-TFT-88-34",
    doi = "10.1016/0550-3213(89)90015-1",
    journal = "Nucl. Phys. B",
    volume = "320",
    pages = "625--668",
    year = "1989"
}

@article{Altschuler:1987zb,
    author = "Altschuler, Daniel and Bardakci, Korkut and Rabinovici, Eliezer",
    title = "{A Construction of the $c < 1$ Modular Invariant Partition Functions}",
    reportNumber = "UCB-PTH-87-47, LBL-24174",
    doi = "10.1007/BF01218579",
    journal = "Commun. Math. Phys.",
    volume = "118",
    pages = "241",
    year = "1988"
}

@article{Goddard:1986ee,
    author = "Goddard, P. and Kent, A. and Olive, David I.",
    title = "{Unitary Representations of the Virasoro and Super Virasoro Algebras}",
    reportNumber = "DAMTP-85-21",
    doi = "10.1007/BF01464283",
    journal = "Commun. Math. Phys.",
    volume = "103",
    pages = "105--119",
    year = "1986"
}

@article{Bardakci:1987ee,
    author = "Bardakci, K. and Rabinovici, E. and Saering, B.",
    title = "{String Models with c \ensuremath{<} 1 Components}",
    reportNumber = "CERN-TH-4760-87",
    doi = "10.1016/0550-3213(88)90470-1",
    journal = "Nucl. Phys. B",
    volume = "299",
    pages = "151",
    year = "1988"
}

@article{Axelrod:1989xt,
    author = "Axelrod, Scott and Della Pietra, Steve and Witten, Edward",
    title = "{Geometric quantization of Chern-Simons gauge theory}",
    reportNumber = "IASSNS-HEP-89/57",
    journal = "J. Diff. Geom.",
    volume = "33",
    number = "3",
    pages = "787--902",
    year = "1991"
}

@Inbook{Gawedzki:1987ak,
author="Gaw{\k{e}}dzki, K.",
editor="'t Hooft, G.
and Jaffe, A.
and Mack, G.
and Mitter, P. K.
and Stora, R.",
title="Topological Actions in Two-Dimensional Quantum Field Thories",
bookTitle="Nonperturbative Quantum Field Theory",
year="1988",
publisher="Springer US",
address="New York, NY",
pages="101--141",
isbn="978-1-4613-0729-7",
doi="10.1007/978-1-4613-0729-7_5",
url="https://doi.org/10.1007/978-1-4613-0729-7_5"
}

@article{Felder:1988sd,
    author = "Felder, Giovanni and Gawedzki, K. and Kupiainen, A.",
    title = "{Spectra of {Wess-Zumino-Witten} Models With Arbitrary Simple Groups}",
    doi = "10.1007/BF01228414",
    journal = "Commun. Math. Phys.",
    volume = "117",
    pages = "127--158",
    year = "1988"
}

@book{DiFrancesco:1997nk,
    author = "Di Francesco, P. and Mathieu, P. and Senechal, D.",
    title = "{Conformal Field Theory}",
    doi = "10.1007/978-1-4612-2256-9",
    isbn = "978-0-387-94785-3, 978-1-4612-7475-9",
    publisher = "Springer-Verlag",
    address = "New York",
    series = "Graduate Texts in Contemporary Physics",
    year = "1997"
}

@inproceedings{Moore:2003vf,
    author = "Moore, Gregory W.",
    title = "{K theory from a physical perspective}",
    booktitle = "{Symposium on Topology, Geometry and Quantum Field Theory (Segalfest)}",
    eprint = "hep-th/0304018",
    archivePrefix = "arXiv",
    reportNumber = "RUNHETC-2003-02",
    pages = "194--234",
    month = "4",
    year = "2003"
}

@book{lepowsky2012introduction,
  title={Introduction to Vertex Operator Algebras and Their Representations},
  author={Lepowsky, J. and Li, H.},
  isbn={9780817681869},
  lccn={2003063839},
  series={Progress in Mathematics},
  url={https://books.google.com/books?id=bB3UBwAAQBAJ},
  year={2012},
  publisher={Birkh{\"a}user Boston}
}

@article{10.1215/S0012-7094-92-06604-X,
author = {Igor B. Frenkel and Yongchang Zhu},
title = {{Vertex operator algebras associated to representations of affine and Virasoro algebras}},
volume = {66},
journal = {Duke Mathematical Journal},
number = {1},
publisher = {Duke University Press},
pages = {123 -- 168},
year = {1992},
doi = {10.1215/S0012-7094-92-06604-X},
URL = {https://doi.org/10.1215/S0012-7094-92-06604-X}
}

@book{frenkel1993axiomatic,
  title={On Axiomatic Approaches to Vertex Operator Algebras and Modules},
  author={Frenkel, I. and Huang, Y.Z. and Lepowsky, J.},
  isbn={9780821825556},
  lccn={lc93017169},
  series={American Mathematical Society: Memoirs of the American Mathematical Society},
  url={https://books.google.com/books?id=gYHUCQAAQBAJ},
  year={1993},
  publisher={American Mathematical Society}
}

@article{dong1995regularity,
  title={Regularity of rational vertex operator algebras},
  author={Dong, Chongying and Li, Haisheng and Mason, Geoffrey},
  journal={arXiv preprint q-alg/9508018},
  year={1995}
}

@article{Cordova:2023jip,
    author = "Cordova, Clay and Garc{\'\i}a-Sep{\'u}lveda, Diego",
    title = "{Non-Invertible Anyon Condensation and Level-Rank Dualities}",
    eprint = "2312.16317",
    archivePrefix = "arXiv",
    primaryClass = "hep-th",
    month = "12",
    year = "2023"
}

@article{rabinovici1988aspects,
  title={Aspects of a {Lagrangian} formulation for modular invariant coset constructions},
  author={Rabinovici, Eliezer},
  journal={Nuclear Physics B-Proceedings Supplements},
  volume={5},
  number={2},
  pages={192--198},
  year={1988},
  publisher={Elsevier}
}

@article{Schellekens:1986mb,
    author = "Schellekens, A. N. and Warner, N. P.",
    title = "{Conformal Subalgebras of {Kac-Moody} Algebras}",
    reportNumber = "CERN-TH-4405/86",
    doi = "10.1103/PhysRevD.34.3092",
    journal = "Phys. Rev. D",
    volume = "34",
    pages = "3092",
    year = "1986"
}

@article{Freed:2022qnc,
    author = "Freed, Daniel S. and Moore, Gregory W. and Teleman, Constantin",
    title = "{Topological symmetry in quantum field theory}",
    eprint = "2209.07471",
    archivePrefix = "arXiv",
    primaryClass = "hep-th",
    month = "9",
    year = "2022"
}

@article{Hori:1994nc,
    author = "Hori, Kentaro",
    title = "{Global aspects of gauged Wess-Zumino-Witten models}",
    eprint = "hep-th/9411134",
    archivePrefix = "arXiv",
    reportNumber = "UT-685, UT-KOMABA-94-1",
    doi = "10.1007/BF02506384",
    journal = "Commun. Math. Phys.",
    volume = "182",
    pages = "1--32",
    year = "1996"
}

@article{Reshetikhin:1991tc,
    author = "Reshetikhin, N. and Turaev, V. G.",
    title = "{Invariants of three manifolds via link polynomials and quantum groups}",
    doi = "10.1007/BF01239527",
    journal = "Invent. Math.",
    volume = "103",
    pages = "547--597",
    year = "1991"
}

@article{Moore:1991ks,
    author = "Moore, Gregory W. and Read, N.",
    title = "{Nonabelions in the fractional quantum Hall effect}",
    doi = "10.1016/0550-3213(91)90407-O",
    journal = "Nucl. Phys. B",
    volume = "360",
    pages = "362--396",
    year = "1991"
}

@article{Kitaev:2005hzj,
    author = "Kitaev, Alexei",
    title = "{Anyons in an exactly solved model and beyond}",
    eprint = "cond-mat/0506438",
    archivePrefix = "arXiv",
    doi = "10.1016/j.aop.2005.10.005",
    journal = "Annals Phys.",
    volume = "321",
    number = "1",
    pages = "2--111",
    year = "2006"
}

@inproceedings{Freed:2009qp,
    author = "Freed, Daniel S. and Hopkins, Michael J. and Lurie, Jacob and Teleman, Constantin",
    title = "{Topological Quantum Field Theories from Compact Lie Groups}",
    booktitle = "{A Celebration of Raoul Bott's Legacy in Mathematics}",
    eprint = "0905.0731",
    archivePrefix = "arXiv",
    primaryClass = "math.AT",
    reportNumber = "AIM-2009-21",
    month = "5",
    year = "2009"
}

@article{Schafer-Nameki:2003nzb,
    author = "Schafer-Nameki, Sakura",
    title = "{D-branes in N=2 coset models and twisted equivariant K theory}",
    eprint = "hep-th/0308058",
    archivePrefix = "arXiv",
    reportNumber = "DAMTP-2003-76",
    month = "8",
    year = "2003"
}

@article{Hellerman:2006zs,
    author = "Hellerman, Simeon and Henriques, Andre and Pantev, Tony and Sharpe, Eric and Ando, Matt",
    title = "{Cluster decomposition, T-duality, and gerby CFT's}",
    eprint = "hep-th/0606034",
    archivePrefix = "arXiv",
    doi = "10.4310/ATMP.2007.v11.n5.a2",
    journal = "Adv. Theor. Math. Phys.",
    volume = "11",
    number = "5",
    pages = "751--818",
    year = "2007"
}

@article{Pantev:2005rh,
    author = "Pantev, Tony and Sharpe, Eric",
    title = "{Notes on gauging noneffective group actions}",
    eprint = "hep-th/0502027",
    archivePrefix = "arXiv",
    month = "2",
    year = "2005"
}

@article{Pantev:2022pbf,
    author = "Pantev, T. and Sharpe, E.",
    title = "{Decomposition in Chern-Simons theories in three dimensions}",
    eprint = "2206.14824",
    archivePrefix = "arXiv",
    primaryClass = "hep-th",
    doi = "10.1142/S0217751X2250227X",
    journal = "Int. J. Mod. Phys. A",
    volume = "37",
    pages = "2250227",
    year = "2022"
}

@article{Nguyen:2021naa,
    author = {Nguyen, Mendel and Tanizaki, Yuya and {\"U}nsal, Mithat},
    title = "{Noninvertible 1-form symmetry and Casimir scaling in 2D Yang-Mills theory}",
    eprint = "2104.01824",
    archivePrefix = "arXiv",
    primaryClass = "hep-th",
    reportNumber = "YITP-21-29",
    doi = "10.1103/PhysRevD.104.065003",
    journal = "Phys. Rev. D",
    volume = "104",
    number = "6",
    pages = "065003",
    year = "2021"
}

@article{Cherman:2019hbq,
    author = {Cherman, Aleksey and Jacobson, Theodore and Tanizaki, Yuya and {\"U}nsal, Mithat},
    title = "{Anomalies, a mod 2 index, and dynamics of 2d adjoint QCD}",
    eprint = "1908.09858",
    archivePrefix = "arXiv",
    primaryClass = "hep-th",
    doi = "10.21468/SciPostPhys.8.5.072",
    journal = "SciPost Phys.",
    volume = "8",
    number = "5",
    pages = "072",
    year = "2020"
}

@article{Huang:2021zvu,
    author = "Huang, Tzu-Chen and Lin, Ying-Hsuan and Seifnashri, Sahand",
    title = "{Construction of two-dimensional topological field theories with non-invertible symmetries}",
    eprint = "2110.02958",
    archivePrefix = "arXiv",
    primaryClass = "hep-th",
    reportNumber = "CALT-TH-2021-035, YITP-SB-2021-19",
    doi = "10.1007/JHEP12(2021)028",
    journal = "JHEP",
    volume = "12",
    pages = "028",
    year = "2021"
}

@article{Pantev:2005zs,
    author = "Pantev, Tony and Sharpe, Eric",
    title = "{GLSM's for Gerbes (and other toric stacks)}",
    eprint = "hep-th/0502053",
    archivePrefix = "arXiv",
    doi = "10.4310/ATMP.2006.v10.n1.a4",
    journal = "Adv. Theor. Math. Phys.",
    volume = "10",
    number = "1",
    pages = "77--121",
    year = "2006"
}

@article{Rocek:1991ps,
    author = "Rocek, Martin and Verlinde, Erik P.",
    title = "{Duality, quotients, and currents}",
    eprint = "hep-th/9110053",
    archivePrefix = "arXiv",
    reportNumber = "IASSNS-HEP-91-68, ITP-SB-91-53",
    doi = "10.1016/0550-3213(92)90269-H",
    journal = "Nucl. Phys. B",
    volume = "373",
    pages = "630--646",
    year = "1992"
}

@inbook{Sharpe:2022ene,
    author = "Sharpe, Eric",
    title = "{An introduction to decomposition}",
    eprint = "2204.09117",
    archivePrefix = "arXiv",
    primaryClass = "hep-th",
    doi = "10.1007/978-3-031-47417-0_8",
    year = "2024"
}

@article{Turaev:1999yf,
    author = "Turaev, Vladimir",
    title = "{Homotopy field theory in dimension two and group algebras}",
    eprint = "math/9910010",
    archivePrefix = "arXiv",
    month = "10",
    year = "1999"
}

@article{Freed:2003qx,
    author = "Freed, Daniel S. and Hopkins, Michael J. and Teleman, Constantin",
    title = "{Twisted K-theory and loop group representations. 1.}",
    eprint = "math/0312155",
    archivePrefix = "arXiv",
    month = "11",
    year = "2003"
}

@unpublished{MRS,
  author = {Moore, Gregory and Rabinovici, Eliezer and Singh, R. K.},
  title  = {Gauged {WZW} models vs. the {GKO} construction},
  note   = {In preparation},
  year   = {2026}
}

@article{Elitzur:2002rt,
    author = "Elitzur, S. and Giveon, A. and Kutasov, D. and Rabinovici, E.",
    title = "{From big bang to big crunch and beyond}",
    eprint = "hep-th/0204189",
    archivePrefix = "arXiv",
    reportNumber = "EFI-02-72, RI-04-02",
    doi = "10.1088/1126-6708/2002/06/017",
    journal = "JHEP",
    volume = "06",
    pages = "017",
    year = "2002"
}

@article{Nappi:1992kv,
    author = "Nappi, Chiara R. and Witten, Edward",
    title = "{A Closed, expanding universe in string theory}",
    eprint = "hep-th/9206078",
    archivePrefix = "arXiv",
    reportNumber = "IASSNS-HEP-92-38",
    doi = "10.1016/0370-2693(92)90888-B",
    journal = "Phys. Lett. B",
    volume = "293",
    pages = "309--314",
    year = "1992"
}

@article{Kounnas:1992wc,
    author = "Kounnas, Costas and Lust, Dieter",
    title = "{Cosmological string backgrounds from gauged WZW models}",
    eprint = "hep-th/9205046",
    archivePrefix = "arXiv",
    reportNumber = "CERN-TH-6494-92, LPTENS-92-16",
    doi = "10.1016/0370-2693(92)91361-C",
    journal = "Phys. Lett. B",
    volume = "289",
    pages = "56--60",
    year = "1992"
}

@article{Cordova:2025zkz,
    author = "Cordova, Clay and Garc{\'\i}a-Sep{\'u}lveda, Diego and Harvey, Jeffrey A.",
    title = "{Generalized Level-Rank Duality, Holomorphic Conformal Field Theory, and Non-Invertible Anyon Condensation}",
    eprint = "2512.24419",
    archivePrefix = "arXiv",
    primaryClass = "hep-th",
    month = "12",
    year = "2025"
}

@article{Alvarez-Gaume:1986nqf,
    author = "Alvarez-Gaume, Luis and Moore, Gregory W. and Nelson, Philip C. and Vafa, Cumrun and Bost, J. b.",
    editor = "Stone, M.",
    title = "{Bosonization in Arbitrary Genus}",
    reportNumber = "HUTP-86-A039",
    doi = "10.1016/0370-2693(86)90466-1",
    journal = "Phys. Lett. B",
    volume = "178",
    pages = "41--47",
    year = "1986"
}

@article{Alvarez-Gaume:1987wwg,
    author = "Alvarez-Gaume, Luis and Bost, J. B. and Moore, Gregory W. and Nelson, Philip C. and Vafa, Cumrun",
    editor = "Stone, M.",
    title = "{Bosonization on Higher Genus Riemann Surfaces}",
    reportNumber = "HUTP-86/A062, CERN-TH-4660/87",
    doi = "10.1007/BF01218489",
    journal = "Commun. Math. Phys.",
    volume = "112",
    pages = "503",
    year = "1987"
}

@article{Friedan:1983xq,
    author = "Friedan, Daniel and Qiu, Zong-an and Shenker, Stephen H.",
    title = "{Conformal Invariance, Unitarity and Two-Dimensional Critical Exponents}",
    reportNumber = "EFI-83-66-CHICAGO",
    doi = "10.1103/PhysRevLett.52.1575",
    journal = "Phys. Rev. Lett.",
    volume = "52",
    pages = "1575--1578",
    year = "1984"
}

@article{Giveon:1994fu,
    author = "Giveon, Amit and Porrati, Massimo and Rabinovici, Eliezer",
    title = "{Target space duality in string theory}",
    eprint = "hep-th/9401139",
    archivePrefix = "arXiv",
    reportNumber = "RI-1-94, NYU-TH-94-01-01",
    doi = "10.1016/0370-1573(94)90070-1",
    journal = "Phys. Rept.",
    volume = "244",
    pages = "77--202",
    year = "1994"
}

@article{Giveon:2003gb,
    author = "Giveon, Amit and Rabinovici, Eliezer and Sever, Amit",
    editor = "Lust, D. and Dorn, Harald",
    title = "{Strings in singular time dependent backgrounds}",
    eprint = "hep-th/0305137",
    archivePrefix = "arXiv",
    reportNumber = "RI-01-03",
    doi = "10.1002/prop.200310102",
    journal = "Fortsch. Phys.",
    volume = "51",
    pages = "805--823",
    year = "2003"
}

@article{Belavin:1984vu,
    author = "Belavin, A. A. and Polyakov, Alexander M. and Zamolodchikov, A. B.",
    editor = "Khalatnikov, I. M. and Mineev, V. P.",
    title = "{Infinite Conformal Symmetry in Two-Dimensional Quantum Field Theory}",
    reportNumber = "CERN-TH-3827",
    doi = "10.1016/0550-3213(84)90052-X",
    journal = "Nucl. Phys. B",
    volume = "241",
    pages = "333--380",
    year = "1984"
}

@article{Freed:2020qfy,
    author = "Freed, Daniel S. and Teleman, Constantin",
    title = "{Gapped Boundary Theories in Three Dimensions}",
    eprint = "2006.10200",
    archivePrefix = "arXiv",
    primaryClass = "math.QA",
    doi = "10.1007/s00220-021-04192-x",
    journal = "Commun. Math. Phys.",
    volume = "388",
    number = "2",
    pages = "845--892",
    year = "2021"
}

@article{Elitzur:1986ye,
    author = "Elitzur, S. and Gross, E. and Rabinovici, E. and Seiberg, N.",
    title = "{Aspects of Bosonization in String Theory}",
    reportNumber = "Print-86-0950 (WEIZMANN), WIS-86/32a-Ph",
    doi = "10.1016/0550-3213(87)90281-1",
    journal = "Nucl. Phys. B",
    volume = "283",
    pages = "413--432",
    year = "1987"
}

@article{Delmastro:2021otj,
    author = "Delmastro, Diego and Gomis, Jaume and Yu, Matthew",
    title = "{Infrared phases of 2d QCD}",
    eprint = "2108.02202",
    archivePrefix = "arXiv",
    primaryClass = "hep-th",
    doi = "10.1007/JHEP02(2023)157",
    journal = "JHEP",
    volume = "02",
    pages = "157",
    year = "2023"
}

@article{Figueroa-OFarrill:1994uwr,
    author = "Figueroa-O'Farrill, Jose M. and Stanciu, Sonia",
    title = "{Equivariant cohomology and gauged bosonic sigma models}",
    eprint = "hep-th/9407149",
    archivePrefix = "arXiv",
    reportNumber = "QMW-PH-94-17",
    month = "7",
    year = "1994"
}

@article{Figueroa-OFarrill:1994vwl,
    author = "Figueroa-O'Farrill, Jose M. and Stanciu, Sonia",
    title = "{Gauged Wess-Zumino terms and equivariant cohomology}",
    eprint = "hep-th/9407196",
    archivePrefix = "arXiv",
    reportNumber = "QMW-PH-94-22",
    doi = "10.1016/0370-2693(94)90304-2",
    journal = "Phys. Lett. B",
    volume = "341",
    pages = "153--159",
    year = "1994"
}

@article{Freed:2006mx,
    author = "Freed, Daniel S.",
    title = "{Pions and Generalized Cohomology}",
    eprint = "hep-th/0607134",
    archivePrefix = "arXiv",
    journal = "J. Diff. Geom.",
    volume = "80",
    number = "1",
    pages = "45--77",
    year = "2008"
}

@article{Kapustin:2010if,
    author = "Kapustin, Anton and Saulina, Natalia",
    editor = "Sati, Hisham and Schreiber, Urs",
    title = "{Surface operators in 3d Topological Field Theory and 2d Rational Conformal Field Theory}",
    eprint = "1012.0911",
    archivePrefix = "arXiv",
    primaryClass = "hep-th",
    reportNumber = "PI-STRINGS-195",
    pages = "175--198",
    month = "12",
    year = "2010"
}

@article{Frohlich:2003hm,
    author = "Frohlich, Jurg and Fuchs, Jurgen and Runkel, Ingo and Schweigert, Christoph",
    title = "{Correspondences of ribbon categories}",
    eprint = "math/0309465",
    archivePrefix = "arXiv",
    reportNumber = "HU-EP-03-31",
    doi = "10.1016/j.aim.2005.04.007",
    journal = "Adv. Math.",
    volume = "199",
    pages = "192--329",
    year = "2006"
}

@article{kirillov2002q,
  title={On a q-analogue of the McKay correspondence and the ADE classification of sl2 conformal field theories},
  author={Kirillov Jr, Alexander and Ostrik, Viktor},
  journal={Advances in Mathematics},
  volume={171},
  number={2},
  pages={183--227},
  year={2002},
  publisher={Elsevier}
}

@book{GelfandVilenkin1964,
  author    = {I. M. Gel'fand and N. Ya. Vilenkin},
  title     = {Generalized Functions, Volume 4: Applications of Harmonic Analysis},
  publisher = {Academic Press},
  year      = {1964},
  address   = {New York}
}

@book{GelfandShilov1964,
  author    = {I. M. Gel'fand and G. E. Shilov},
  title     = {Generalized Functions, Volume 1: Properties and Operations},
  publisher = {Academic Press},
  year      = {1964},
  address   = {New York}
}

@article{huang2015braided,
  title={Braided tensor categories and extensions of vertex operator algebras},
  author={Huang, Yi-Zhi and Kirillov Jr, Alexander and Lepowsky, James},
  journal={Communications in Mathematical Physics},
  volume={337},
  number={3},
  pages={1143--1159},
  year={2015},
  publisher={Springer}
}

@book{turaev1994quantum,
  title={Quantum Invariants of Knots and 3-Manifolds},
  author={Turaev, Vladimir G},
  year={1994},
  publisher={Walter de Gruyter},
  address={Berlin, New York},
  series={De Gruyter Studies in Mathematics},
  volume={18},
  isbn={9783110137040}
}

@article{bais2009condensate,
  title={Condensate-induced transitions between topologically ordered phases},
  author={Bais, F Alexander and Slingerland, JK},
  journal={Physical Review B—Condensed Matter and Materials Physics},
  volume={79},
  number={4},
  pages={045316},
  year={2009},
  publisher={APS}
}

@article{kong2014anyon,
  title={Anyon condensation and tensor categories},
  author={Kong, Liang},
  journal={Nuclear Physics B},
  volume={886},
  pages={436--482},
  year={2014},
  publisher={Elsevier}
}

@article{Schellekens:1990xy,
    author = "Schellekens, A. N. and Yankielowicz, S.",
    title = "{Simple Currents, Modular Invariants and Fixed Points}",
    reportNumber = "CERN-TH-5622-90",
    doi = "10.1142/S0217751X90001367",
    journal = "Int. J. Mod. Phys. A",
    volume = "5",
    pages = "2903--2952",
    year = "1990"
}

@article{Frohlich:2003hg,
    author = "Frohlich, Jurg and Fuchs, Jurgen and Runkel, Ingo and Schweigert, Christoph",
    editor = "Dorn, Harald and Lust, D.",
    title = "{Algebras in tensor categories and coset conformal field theories}",
    eprint = "hep-th/0309269",
    archivePrefix = "arXiv",
    reportNumber = "HU-EP-03-68",
    doi = "10.1002/prop.200310162",
    journal = "Fortsch. Phys.",
    volume = "52",
    pages = "672--677",
    year = "2004"
}

@article{turaev1992modular,
  title={Modular categories and 3-manifold invariants},
  author={Turaev, Vladimir G},
  journal={International Journal of Modern Physics B},
  volume={6},
  number={11n12},
  pages={1807--1824},
  year={1992},
  publisher={World Scientific}
}

@article{Schellekens:1989uf,
    author = "Schellekens, A. N. and Yankielowicz, S.",
    title = "{Field Identification Fixed Points in the Coset Construction}",
    reportNumber = "CERN-TH-5483-89",
    doi = "10.1016/0550-3213(90)90657-Y",
    journal = "Nucl. Phys. B",
    volume = "334",
    pages = "67--102",
    year = "1990"
}

@article{Cordova:2024goh,
    author = "Cordova, Clay and Garc{\'\i}a-Sep{\'u}lveda, Diego",
    title = "{Topological Cosets via Anyon Condensation and Applications to Gapped $\mathrm{\bf{QCD_{2}}}$}",
    eprint = "2412.01877",
    archivePrefix = "arXiv",
    primaryClass = "hep-th",
    month = "12",
    year = "2024"
}

@article{Cardy:1989ir,
    author = "Cardy, John L.",
    title = "{Boundary Conditions, Fusion Rules and the Verlinde Formula}",
    reportNumber = "UCSB-TH-89-06",
    doi = "10.1016/0550-3213(89)90521-X",
    journal = "Nucl. Phys. B",
    volume = "324",
    pages = "581--596",
    year = "1989"
}

@article{Freed:2007wja,
    author = "Freed, Daniel S. and Hopkins, Michael J. and Teleman, Constantin",
    title = "{Loop groups and twisted K-theory I}",
    eprint = "0711.1906",
    archivePrefix = "arXiv",
    primaryClass = "math.AT",
    doi = "10.1112/jtopol/jtr019",
    journal = "J. Topol.",
    volume = "4",
    number = "4",
    pages = "737--798",
    year = "2011"
}

@article{Witten:1991yr,
    author = "Witten, Edward",
    title = "{On string theory and black holes}",
    reportNumber = "IASSNS-HEP-91-12",
    doi = "10.1103/PhysRevD.44.314",
    journal = "Phys. Rev. D",
    volume = "44",
    pages = "314--324",
    year = "1991"
}

@article{SPIEGELGLAS199036,
title = {Spin sums, fusion rules and correlators by filling},
journal = {Physics Letters B},
volume = {247},
number = {1},
pages = {36-40},
year = {1990},
issn = {0370-2693},
doi = {https://doi.org/10.1016/0370-2693(90)91045-D},
url = {https://www.sciencedirect.com/science/article/pii/037026939091045D},
author = {Mordechai Spiegelglas}
}

@article{Spiegelglas:1991uc,
    author = "Spiegelglas, Mordechai",
    title = "{Setting fusion rings in topological Landau-Ginzburg}",
    reportNumber = "TECHNION-PH-8-91",
    doi = "10.1016/0370-2693(92)90298-I",
    journal = "Phys. Lett. B",
    volume = "274",
    pages = "21--26",
    year = "1992"
}

@article{Spiegelglas:1992jg,
    author = "Spiegelglas, Mordechai and Yankielowicz, Shimon",
    title = "{G/G topological field theories by cosetting G(k)}",
    eprint = "hep-th/9201036",
    archivePrefix = "arXiv",
    reportNumber = "TECHNION-PH-90-34, TAUP-1934-91",
    doi = "10.1016/0550-3213(93)90247-M",
    journal = "Nucl. Phys. B",
    volume = "393",
    pages = "301--336",
    year = "1993"
}

@article{Ishibashi:1988kg,
    author = "Ishibashi, Nobuyuki",
    title = "{The Boundary and Crosscap States in Conformal Field Theories}",
    reportNumber = "UT-530-TOKYO",
    doi = "10.1142/S0217732389000320",
    journal = "Mod. Phys. Lett. A",
    volume = "4",
    pages = "251",
    year = "1989"
}

@article{Mandal:1991tz,
    author = "Mandal, Gautam and Sengupta, Anirvan M. and Wadia, Spenta R.",
    title = "{Classical solutions of two-dimensional string theory}",
    reportNumber = "IASSNS-HEP-91-10",
    doi = "10.1142/S0217732391001822",
    journal = "Mod. Phys. Lett. A",
    volume = "6",
    pages = "1685--1692",
    year = "1991"
}

@article{Moore:2006dw,
    author = "Moore, Gregory W. and Segal, Graeme",
    title = "{D-branes and K-theory in 2D topological field theory}",
    eprint = "hep-th/0609042",
    archivePrefix = "arXiv",
    reportNumber = "RUNHETC-06-12",
    month = "8",
    year = "2006"
}
\end{document}